\documentclass[usenatbib,usea4paper,fleqn]{mn2e}
\usepackage{amssymb}
\usepackage{color}
\usepackage{url}
\usepackage{aas_macros}
\usepackage{graphicx}
\usepackage{times}
\usepackage[T1]{fontenc}
\usepackage{aecompl,amsmath}

\def \mathbi#1{\textbf{\em #1}}

\date{}
\pagerange{\pageref{firstpage}--\pageref{lastpage}}
\pubyear{2014}

\begin{document}
\label{firstpage}

\title[On the Bardeen-Petterson effect]{On the Bardeen-Petterson effect in black hole accretion discs}
\defcitealias{lubow2002evolution}{LOP02}
\author[Nealon, Price \& Nixon]
{Rebecca Nealon$^{1}$\thanks{rebecca.nealon@monash.edu}, 
Daniel J. Price$^1$\thanks{daniel.price@monash.edu} \& 
Chris J. Nixon$^2$\thanks{chris.nixon@jila.colorado.edu}\thanks{Einstein Fellow}\\ 
$^1$ Monash Centre for Astrophysics and School of Mathematical Sciences, Monash University, Vic 3800, Australia\\
$^2$ JILA, University of Colorado \& NIST, Boulder CO 80309-0440, USA
} 
\maketitle

\begin{abstract}
We investigate the effect of black hole spin on warped or misaligned accretion discs --- in particular i) whether or not the inner disc edge aligns with the black hole spin and ii) whether the disc can maintain a smooth transition between an aligned inner disc and a misaligned outer disc, known as the Bardeen-Petterson effect. We employ high resolution 3D smoothed particle hydrodynamics simulations of $\alpha$-discs subject to Lense-Thirring precession, focussing on the bending wave regime where the disc viscosity is smaller than the aspect ratio $\alpha \lesssim H/R$. We first address the controversy in the literature regarding possible steady-state oscillations of the tilt close to the black hole. We successfully recover such oscillations in 3D at both small and moderate inclinations ($\lesssim15^{\circ}$), provided both Lense-Thirring and Einstein precession are present, sufficient resolution is employed, and provided the disc is not so thick so as to simply accrete misaligned. Second, we find that discs inclined by more than a few degrees in general steepen and break rather than maintain a smooth transition, again in contrast to previous findings, but only once the disc scale height is adequately resolved. Finally, we find that when the disc plane is misaligned to the black hole spin by a large angle, the disc `tears' into discrete rings which precess effectively independently and cause rapid accretion, consistent with previous findings in the diffusive regime ($\alpha \gtrsim H/R$). Thus misalignment between the disc and the spin axis of the black hole provides a robust mechanism for growing black holes quickly, regardless of whether the disc is thick or thin.

\end{abstract}


\section{Introduction}
\label{section:intro}
\citet{bardeen_petterson_1975} first computed the evolution of a warped accretion disc subjected to Lense-Thirring precession \citep{Lense:1918vn} caused by frame-dragging from the spin of a central black hole. Although their original equations were later shown to be incorrect \citep{pap_pringle_1983}, their qualitative findings of an aligned inner disc smoothly connected to a misaligned outer disc --- the `Bardeen-Petterson effect' --- has been confirmed both from subsequent 1D calculations with corrected equations \citep{Kumar:1985qy,pringle_1992} and in three dimensional smoothed particle hydrodynamics (SPH) simulations by \citet{nelson_pap_2000}.

%

 \citet{pap_pringle_1983} showed that there were two regimes for warp propagation in discs depending on the ratio of the effective viscosity $\alpha$ to the aspect ratio $H/R$. For $\alpha \gtrsim H/R$, warps can be described with a diffusion equation, whereas for $\alpha \lesssim H/R$ they propagate as bending waves at half the sound speed \citep{Papaloizou:1995pn}. Previous studies by \citet{Kumar:1985qy} and \citet{pringle_1992} were performed in the diffusive regime. The first studies performed in the bending wave regime \citep*[][hereafter LOP02]{i_and_i_1997,lubow2002evolution} suggested a conflict with the Bardeen-Petterson picture --- finding that the black hole spin could drive the disc tilt into a steady state that is oscillatory and non-zero. This implies that the inner edge of the disc may be misaligned with respect to the black hole spin. As simple pictures of gas accretion favour alignment of the outer accretion disc with the galaxy disc, this has been suggested as an explanation for the observed random orientation of jets with respect to their host galaxies (\citealt{kinney_2000}; \citetalias{lubow2002evolution}). However, unlike the diffusive regime for which there now exists a full non-linear theory describing warps of arbitrary amplitude and $\alpha$ \citep{Ogilvie:1999lr,Ogilvie:2000uq} only linear theory exists for the bending wave regime (\citealt{Papaloizou:1995pn,lubow_ogilvie_2000}; \citetalias{lubow2002evolution}; though see \citealt{ogilvie_2006}); hence these studies apply only to small amplitude warps and could not account for non-linear effects.


The first 3D numerical simulations of the Bardeen-Petterson effect were performed by \citet{nelson_pap_2000}, using a post-Newtonian description of the central potential. Their simulations in both the wavelike and diffusive regimes largely confirmed the Bardeen-Petterson effect, namely an aligned inner edge, a smooth transition to the outer disc and importantly, no evidence of oscillations in the tilt. The reason for the discrepancy with \citetalias{lubow2002evolution} is unclear, with \citet{nelson_pap_2000} claiming that `non-linear effects lead to the damping of these short wavelength features' whilst \citetalias{lubow2002evolution} show that these features are not small compared to the local disc scale height. Further complicating matters, recent simulations that included a full general relativistic (GR) treatment may have found these oscillations and a non-zero tilt at the inner edge \citep{Fragile:2007uq,Zhuravlev:2014fk}.

A further challenge to the Bardeen-Petterson description has arisen from recent simulations of warped discs in the diffusive regime ($\alpha \gtrsim H/R$). 3D simulations of warps in isolated discs have demonstrated close agreement with the non-linear theory of \citet{Ogilvie:1999lr} \citep[see][]{lodato_2010}. However, in simulations of warps driven by Lense-Thirring precession it was found that instead of maintaining a smooth transition (as described by the linear Bardeen-Petterson effect), discs break when the angle between the disc and the black hole spin is more than a few degrees \citep{nixon_et_al_2012}. Rings of gas were found to be torn off the disc, precessing effectively independently before being accreted. The net effect was a much higher accretion rate on to the black hole. Thus the Bardeen-Petterson idea of a smooth transition does not appear to hold in the diffusive regime for an arbitrary choice of parameters. It is not clear whether or not misaligned discs in the wavelike regime will also break. 

Here we reexamine the Bardeen-Petterson effect in black hole accretion discs. Building on the success of earlier studies we use the SPH code \textsc{Phantom} to model warped discs in 3D \citetext{\citealt{lodato_2010,nixon_2012}; \citealt*{nixon_et_al_2012b}; \citealt{nixon_et_al_2012}; \citealt*{facchini_2013,nixon_2013}; \citealt{Martin:2014aa}, \citealt{Martin:2014bb}}. Here we focus on the wavelike propagation regime ($\alpha \lesssim H/R$). We start by considering the possible reasons for the discrepancy between \citetalias{lubow2002evolution} and \citet{nelson_pap_2000}, as well as two possible complications to the Bardeen-Petterson picture of an aligned inner disc smoothly joined to a misaligned outer disc in Section~\ref{section:warps}. We present the numerical method and tests of wavelike warp propagation with SPH in Section~\ref{sec:method}. In Section~\ref{section:results} we present a suite of 3D simulations designed to confirm under what circumstances the Bardeen-Petterson description holds. In Section~\ref{section:disc} we discuss the implications of these and in Section~\ref{section:conclusion} we outline our conclusions.

An obvious caveat is that we do not consider magnetic fields, even though it is widely accepted that magnetorotational instability (MRI) is the controlling mechanism for viscosity in the disc \citep{balbus_hawley_91}. However, we can still capture the dominant behaviour in the disc as magnetic fields have been shown to have little effect on the geometrical evolution \citep*{Sorathia:2013fk}. We also follow \citet{nelson_pap_2000} in using using a post-Newtonian approach instead of general relativity (GR). As we will see, one of the findings of this paper is that this approximation must be considered carefully in order to capture the combination of relativistic effects that lead to tilt oscillations.

\begin{figure*}
\includegraphics [width=\textwidth] {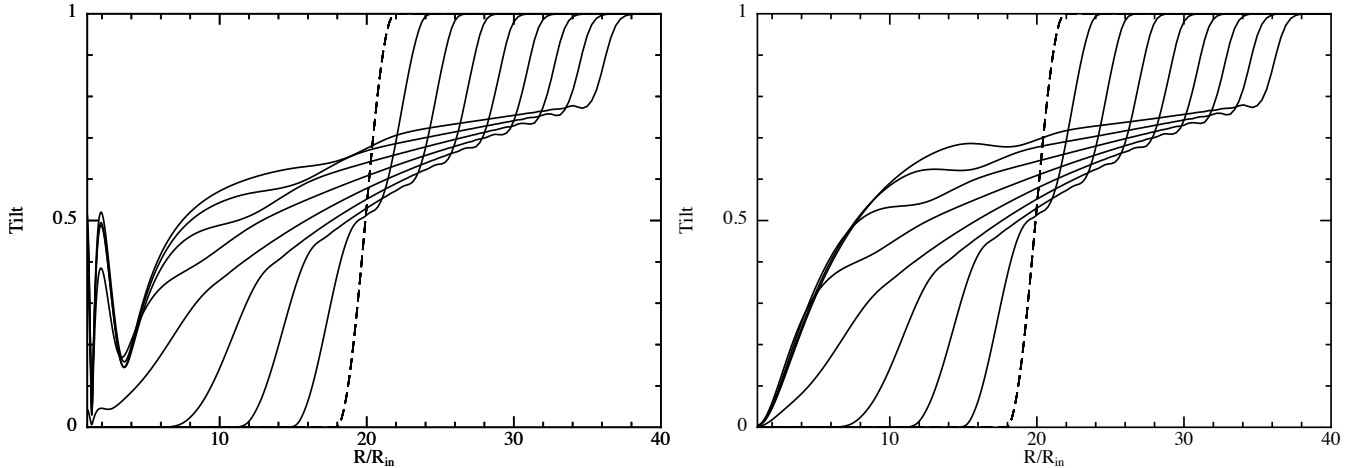}
\caption{The steady state tilt profile found by \protect\citetalias{lubow2002evolution}, as in their Figure 3 and shown to the same time (at eight equally spaced times). The original apsidal and nodal frequencies are used in the left panel (Equations~\ref{equation:apsidal}~and~\ref{equation:nodal}) with the same sign. In the right panel we have reversed the sign of the nodal frequency so the precession frequencies have different signs, and there are no oscillations present in this steady state (similar to \protect\citetalias{lubow2002evolution}, Figure 6). Here the tilt is shown as a fraction of the maximum tilt, the initial condition is shown with the dashed line and $R_{\rm in}=4R_{\rm g}$.}
\label{fig:LOP02_fig3}
\end{figure*}

\section{Does the Bardeen-Petterson effect hold in the wavelike regime?}
\label{section:warps}
We consider three possible ways that the Bardeen-Petterson effect may be violated in wavelike discs. Firstly, radial oscillations in the tilt of the disc may prevent the disc from aligning at the inner edge. Secondly, the smooth transition between aligned and misaligned material may be broken if the disc tears, as has been observed in the diffusive regime \citep{nixon_king_2011,nixon_et_al_2012}. Finally, it may not be possible for the disc to find a steady state if the disc is relatively thick and the viscous time is short.

\subsection{Is the inner disc aligned?}
\citetalias{lubow2002evolution} considered warps in geometrically thin, almost Keplerian discs described by a surface density $\Sigma (R)$ and angular velocity $\Omega(R)$. The scale-height of the disc is given by $H(R) \equiv c_{\rm s}/\Omega$, where $c_{\rm s}(R)$ is the sound speed in the disc. Their description is one dimensional in the sense that the total angular momentum in the disc \mathbi{L} is a function only of the cylindrical radial coordinate, $R$. The disc is then discretised into a series of rings, each described by the orientation of its tilt and twist angle. The tilt angle $\beta$ is measured from the $z$-axis, and if this angle varies with radius the disc is considered to be warped. The twist angle $\gamma$ is measured from an axis that is perpendicular to the $z$-axis, and similarly, if the twist angle varies with radius the disc is twisted. These two angles can be related to the unit angular momentum vector by $\mathbi{ l} = \mathbi{L}/L = ( \cos \gamma \sin \beta, \sin \gamma \sin \beta, \cos \beta)$ \citep{pringle_1996}.

We assume an $\alpha$ disc viscosity where $\nu = \alpha c_{\rm s} H$ \citep{shakura_sunyaev}. For accretion discs with $\alpha \lesssim H/R$, the warp propagates as a dispersive wave \citep{pap_pringle_1983,Papaloizou:1995pn}. Assuming that the disc is nearly Keplerian and not self-gravitating, the equations of motion describing the wave propagation are \citep{lubow_ogilvie_2000,lubow2002evolution}

\begin{equation}
\Sigma R^2 \Omega \frac{\partial \mathbi{l}}{\partial t} = \frac{1}{R} \frac{\partial \mathbi{G}}{\partial R} + \mathbi{T}, \label{equation:wave1}
\end{equation}
\begin{equation}
\frac{\partial \mathbi{G}}{\partial t} - \left(\frac{\Omega^2 - \kappa^2}{2\Omega}\right)\mathbi{l} \times \mathbi{G} + \alpha \Omega \mathbi{G} = \Sigma R^3 \Omega \frac{c_{\rm s}^2}{4} \frac{\partial \mathbi{l}}{\partial R}. \label{equation:wave2}
\end{equation}

Here $\kappa$ is the epicyclic frequency, $\mathbi{G}$ represents the internal horizontal torque in the disc and $\mathbi{T}$ is the external torque per unit area. \citetalias{lubow2002evolution} chose a complex representation where the warp is given by $W=l_x+ i l_y$ and the internal torque as $G=G_x + i G_y$. This allows Equations~\ref{equation:wave1}~and~\ref{equation:wave2} to be rewritten as
\begin{equation}
\Sigma R^2 \Omega \left[ \frac{\partial W}{\partial t} - i \left(\frac{\Omega^2 - \Omega_z^2}{2 \Omega}\right) W \right] = \frac{1}{R} \frac{\partial G}{\partial R}, \label{equation:wave3}\\
\end{equation}
\begin{equation}
\frac{\partial G}{\partial t} - i \left(\frac{\Omega^2 - \kappa^2}{2 \Omega}\right)G + \alpha \Omega G = \frac{P R^3 \Omega}{4} \frac{\partial W}{\partial R}.\label{equation:wave4}
\end{equation}
These equations describe the propagation of a warp in the linear regime, and were solved numerically by \citetalias{lubow2002evolution} to find the steady state shape of the disc around a Kerr black hole. In this case the apsidal and nodal precession frequencies in the disc (scaled by $\Omega$) can be approximated to first order from the Kerr metric as \citep{kato_1990}
\begin{equation}
\eta_{\rm LOP} = \frac{\kappa^2 - \Omega^2}{2 \Omega^2}=-\frac{3}{2}\frac{R_{\rm s}}{R}, \label{equation:apsidal}
\end{equation}
\begin{equation}
\zeta_{\rm LOP} = \frac{\Omega_z^2 - \Omega^2}{2 \Omega^2}=-\frac{a}{\sqrt{2}}\left(\frac{R_{\rm s}}{R}\right)^{3/2}, \label{equation:nodal}
\end{equation}
where $R_{\rm s} = 2GM/c^2$ and $a$ is the black hole spin. These frequencies are used in the solution by inserting them directly into Equations~\ref{equation:wave3}~and~\ref{equation:wave4}. An example of the solution by \citetalias{lubow2002evolution} is shown in the left of Figure~\ref{fig:LOP02_fig3}, with the same parameters used in their work. The steady state solution is formed from the interaction of the ingoing and the outgoing bending waves, where the outgoing waves are created by the reflection of the ingoing waves at the inner boundary \citepalias{lubow2002evolution}. Here the oscillatory behaviour of the steady state near the inner edge is clear, as is the non-zero tilt at the inner edge.

It is known that the relative signs of the apsidal and nodal frequencies determines whether the solution is oscillatory or not \citep{i_and_i_1997}. The frequencies used by \citetalias{lubow2002evolution} have the same sign, leading to radial oscillations in the steady state tilt profile. We confirm that the oscillatory profile is dependent only on the signs of the frequencies by changing the sign of $\zeta_{\rm LOP}$ (equivalent to modelling a retrograde black hole, see \citetalias{lubow2002evolution}, Figure 6)  in the right hand panel of Figure~\ref{fig:LOP02_fig3}. The two solutions evolve in the same manner with the exception of the oscillations near the inner edge.

 While one is free to set the precession frequencies directly when solving Equations \ref{equation:wave3} and \ref{equation:wave4}, in 3D the nodal precession can be induced directly (e.g. using the post-Newtonian description of Lense-Thirring precession from a spinning black hole) and the apsidal precession (i.e. Einstein precession) arises indirectly from the central potential. Hence, it is possible for the choice of potential to preclude oscillations from the steady state solution in 3D simulations. It is then not surprising that simulations that do not take the apsidal precession into account as above also do not report tilt oscillations \citep{Sorathia:2013fk}. However, simulations by \citet{nelson_pap_2000} did make use of a potential that resulted in apsidal and nodal precession frequencies with the same sign but did not find oscillations. Here we use high resolution simulations along with a potential that leads to precession frequencies of the same sign to investigate this discrepancy.

\subsection{When does the disc break?}
\label{subsection:torques}
The derivation of Equations~\ref{equation:wave1}~and~\ref{equation:wave2} assumes that the inclination of the disc is linear. From previous results in the non-linear regime we would anticipate that the disc may break when the external torque applied to the disc is stronger than the internal torque. Here the internal disc communication is governed by a combination of pressure and viscosity. The viscous torque that acts between successive, discrete rings in the disc is given by \citep{Lynden-Bell:1974uq}
\begin{equation}
\mathbi{G} = 3 \pi \nu \Sigma (GMR)^{1/2}.
\label{equation:internal_torque}
\end{equation}
Lense-Thirring precession causes the rings that make up the disc to precess. Per unit area on the disc, this torque is given by \citep[e.g.][]{nixon_et_al_2012}
\begin{equation}
\mathbi{T} = \frac{GM}{2a} \Sigma R^2 \Omega |\sin \theta | \left(\frac{R_{\rm g}}{R}\right)^3,
\end{equation}
where $R_{\rm g} = GM/c^2$, $a$ is the black hole spin and $\theta$ is the angle between the plane of the disc and the direction of the black hole spin. If the external torque applied to the disc is greater than the internal torque maintaining the disc, the rings will precess independently faster than the disc is able to communicate the precession \citep{nixon_et_al_2012}. This will result in the disc being separated and breaking, perhaps into differentially precessing rings. Assuming that the disc has no initial warp and that internal communication is dominated by viscosity, a comparison of the above torques predicts a maximum radius that it is possible for this to occur \citep{nixon_2013}
\begin{equation}
R_{\rm break} \lesssim \left( \frac{4a}{3\alpha} | \sin \theta | \left(\frac{H}{R}\right)^{-1}\right)^{2/3} R_{\rm g}.
\label{equation:breaking_radius}
\end{equation}
This approximate relationship places an upper bound on the breaking radius of the disc at a given angle. For the typical parameters used in this paper, we have $H/R = 0.05$, $R_{\rm out}=40R_{\rm in}$, $\alpha=0.01$ and $a=0.9$. At the outer edge of the disc, the above relation then reduces to
\begin{equation}
R_{\rm break} \lesssim 45 R_{\rm in} \left(\sin \theta \right)^{2/3} R_{\rm g}.
\label{equation:break_radius_est}
\end{equation}

This predicts that tearing may occur in the disc for inclinations of more than $6^{\circ}$. At this inclination or greater one would expect the discs to break rather than align. However, in the bending wave regime that we consider here, the internal communication is dominated by pressure.  In this case we can estimate the radius at which the disc will break by comparing the sound crossing and the precession timescales in the disc. Following \citet{nixon_2013} and assuming that the disc is inviscid (and hence not taking into account any wave damping) we find that
\begin{equation}
R_{\rm break,t} \lesssim \left(4a | \sin \theta | \frac{R}{H} \right)^{2/3} R_{\rm g}.
\label{equation:other_breaking_radius}
\end{equation}
We therefore only expect the disc to break closer to the black hole than Equation~\ref{equation:break_radius_est}.

\subsection{Can the disc accrete misaligned?}
\label{subsection:misaligned}
A further assumption made in developing Equations~\ref{equation:wave1}~and~\ref{equation:wave2} was that the viscous timescale in the disc is much larger than any other timescale, equivalent to assuming that the disc is replenished from radii outside the computational domain, or $\alpha \ll H/R$ \citep{lubow2002evolution}. This implies that the surface density profile does not change during the evolution of these equations, which is valid until the warp reaches the outer boundary. We can quantify this approximation during the evolution of the equations by considering the ratio of the wave and viscous timescales \citep[as in][]{Lodato:2006fk,facchini_2013}
\begin{equation}
\frac{t_{\rm wave}}{t_{\rm \nu}} = \frac{2R\nu}{c_{\rm s} R^2} = 2\alpha \frac{H}{R}.
\end{equation}
For $\alpha < H/R \ll 1$, the above relation implies that the viscous time is much greater than the sound crossing time that the warp communicates, so we can neglect mass accretion. Indeed, \citetalias{lubow2002evolution} neglected the evolution of the surface density profile completely in their solution, equivalent to assuming no mass is accreted at all. However for their disc $H/R=0.1030$, and so it is not clear that this assumption holds. Additionally, the viscous time can be written as
\begin{equation}
t_{\rm \nu} = \frac{1}{\alpha \Omega} \left(\frac{H}{R}\right)^{-2}.
\end{equation}
In this form, it is clear that increasing the aspect ratio of a disc results in a significant decrease in the viscous time. At a given radius $R$, when the viscous timescale is comparable to (or smaller than) the precession timescale at that same radius, accretion dominates. In thick discs (or tori) it may then be possible for the material in the outer disc to be accreted before it has a chance to align. The tilt profile in this case will not reach a steady state but instead be determined by the inward flux of angular momentum. Thus in relatively thick discs no tilt oscillations would be expected.

\section{Numerical Method}
\label{sec:method}
We use 3D simulations to investigate whether the Bardeen-Petterson effect holds in accretion discs focussing on tilt oscillations, large inclinations and misaligned accretion in the limit where $\alpha \lesssim H/R$. We use \textsc{Phantom}, an efficient and low memory SPH code \citep{lodato_2010,price_federrath_2010,Price:2012fj}. This code has been widely used to model warped discs in the diffusive regime and to investigate related problems with warped wavelike discs (see Section~\ref{section:intro}). The physical disc viscosity in the disc is represented in the code using the artificial viscosity ($\alpha_{\rm AV}$) method outlined in \citet{lodato_2010}.

\subsection{Modelling Lense-Thirring Precession}
\label{subsection:lt_prec}
The Lense-Thirring precession exerted by the black hole is modelled using a post-Newtonian approximation, represented as a first order (in $v/c$) correction in the momentum equation. Taking this into account, the momentum equation becomes \citep{nelson_pap_2000}
\begin{equation}
\frac{{\rm d}\mathbi{v}}{{\rm d}t} = -\frac{1}{\rho} \nabla P + \mathbi{v} \times \mathbi{h} - \nabla \Phi + S_{\rm visc}.
\label{equation:momentum}
\end{equation}
Here $S_{\rm visc}$ is the viscous force per unit mass and $\mathbi{v} \times \mathbi{h}$ is the gravomagnetic force per unit mass, with $\mathbi{h}$ given by
\begin{equation}
\mathbi{h} = \frac{2\mathbi{S}}{R^3} - \frac{6(\mathbi{S}\cdot\mathbi{r})\mathbi{r}}{R^5},
\end{equation}
and $\mathbi{S} = aG^2M^2 \mathbi{k}/c^3$, pointing in same direction as the black hole spin.

Assuming a thin disc with linear disturbances, the quantity $\mathbi{v} \times \mathbi{h}$ can be expressed as
\begin{equation}
\mathbi{v} \times \mathbi{h} = \frac{2S\left(\mathbi{v} \times {\hat{\mathbi{k}}}\right)}{R^3} + \frac{6S(z\mathbi{r} \times \mathbi{v})}{R^5}.
\label{equation:vcrossh}
\end{equation}
This expression is mainly useful for setting up the initial orbital velocities for the disc particles (Section~\ref{subsection:scope}).

\subsubsection{Implementation in Leapfrog}
A complication to the implementation in the code is that we use a leapfrog integrator in the `Velocity-Verlet' form, where the positions and velocities of the particles are updated from time $t^n$ to $t^{n+1}$ according to 
\begin{align}
\mathbi{v}^{n+\frac12} & = \mathbi{v}^n + \frac12 \Delta t \mathbi{a}^n, \\
\mathbi{x}^{n+1} & = \mathbi{x}^n + \Delta t \mathbi{v}^{n + \frac12}, \\
\mathbi{v}^{n+1} & = \mathbi{v}^{n+\frac12} + \frac12 \Delta t \mathbi{a}^{n+1}. \label{eq:corr}
\end{align}
However, the acceleration caused by Lense-Thirring precession depends on velocity. Thus  (\ref{eq:corr}) becomes implicit. This can be easily solved by writing the corrector step in the form
\begin{equation}
\mathbi{v}^{n+1} = \mathbi{v}^{n+\frac12} + \frac12 \Delta t \mathbi{a}^{n+1}_{\rm pos} + \frac12 \Delta t  \left(\mathbi{v}^{n+1} \times \mathbi{h}^{n+1}\right),
\end{equation}
where $\mathbi{a}^{n+1}_{\rm pos}$ contains the position-dependent terms. This forms a set of three linear equations for each component of $\mathbi{v}^{n+1}$, that we solve analytically by inverting the resulting 3 x 3 matrix.

\subsubsection{Potentials}
We make use of two gravitational potentials in this work. The first was previously introduced by \citet{nelson_pap_2000}, referred to as the Einstein potential (see their Equation 8). In our notation it is given as
\begin{equation}
\Phi_{\rm E}(R) = -\frac{GM}{R}\left(1+\frac{3R_{\rm g}}{R}\right).
\label{equation:einstein_pot}
\end{equation}
This potential was introduced because it prevents the gravitational force tending to infinity as the radius decreases. However, it also results in the correct apsidal precession frequency at large distances from the black hole and has the same sign as the nodal frequency \citep{nelson_pap_2000}. This is in contrast to the standard Keplerian potential
\begin{equation}
\Phi(R) = -\frac{GM}{R}.
\label{equation:pn_pot}
\end{equation}
The standard potential (\ref{equation:pn_pot}) was used in all of the non-linear simulations, except for Figures~\ref{fig:oscillation}--\ref{fig:matching_NP2000} where (\ref{equation:einstein_pot}) was used.

\subsubsection{Precession Frequencies}
\label{subsubsection:freq}
We calculate the apsidal and nodal precession frequencies in our disc using the standard (Newtonian) definitions for the epicyclic and vertical frequencies,
\begin{equation}
\kappa^2 = 4 \Omega^2 + R\frac{d}{dR}(\Omega^2),
\label{equation:kappa}
\end{equation}
\begin{equation}
\Omega_z^2 \equiv \frac{\partial^2 \Phi(R)}{\partial z^2}\bigg|_{z=0}.
\label{equation:omegaz}
\end{equation}
Following \citet{nelson_pap_2000}, we compute these using an effective potential that takes in to account the second and third terms of Equation~\ref{equation:momentum},
\begin{equation}
\Phi_{\rm eff}(R) = \Phi(R) + \frac{4S\sqrt{GM}}{5R^{5/2}} - \frac{6Sz^2\sqrt{GM}}{R^{9/2}}.
\label{equation:pot}
\end{equation}
This potential accounts for both the normal Keplerian potential and the correction due to the $\bf v \times \bf h$ term, where $\Phi(R)$ could be represented by either Equation~\ref{equation:einstein_pot} or \ref{equation:pn_pot}. Firstly considering the Einstein potential, using Equations~\ref{equation:kappa}~and~\ref{equation:omegaz} the post-Newtonian apsidal and nodal precession frequencies (scaled by $\Omega$) are given by 
\begin{equation}
\eta_{\rm E} = \frac{-1}{2 \Omega^2}\left[\frac{6GMR_g}{R^4} - \frac{3S \sqrt{GM}}{R^{9/2}}\right],
\label{equation:eta_full}
\end{equation}
\begin{equation}
\zeta_{\rm E} = \frac{-2S\sqrt{GM}}{R^{9/2} \Omega^2},
\label{equation:zeta_full}
\end{equation}
where $\Omega^2$ is given by
\begin{equation}
\Omega^2_{\rm E} = \frac{GM}{R^3}\left(1+\frac{6 R_{\rm g}}{R}\right) - \frac{2S \sqrt{GM}}{R^{9/2}}.
\label{equation:omega_z_squared}
\end{equation}
As the signs of the apsidal and nodal precession frequencies here are the same throughout the disc, this potential will allow an oscillatory profile to develop. Considering now a standard post-Newtonian potential, given by Equation~\ref{equation:pn_pot}, we find the apsidal and nodal precession frequencies to be (again scaled by $\Omega$)
\begin{equation}
\eta_{\rm PN} = \frac{3S}{2 \sqrt{GM} R^{3/2} - 4S},
\label{equation:eta}
\end{equation}
\begin{equation}
\zeta_{\rm PN} = \frac{-4S}{2 \sqrt{GM} R^{3/2} - 4S}.
\label{equation:zeta}
\end{equation}
Here we note an important difference with respect to the solution used by \citetalias{lubow2002evolution}. As the signs of the precession frequencies here are opposite, the steady state tilt profile will not have oscillations if the potential in Equation~\ref{equation:pn_pot} is used.

\begin{table}
\begin{center}
\begin{tabular}{l*{4}{c}r}
\hline
Simulation & $\theta$ ($^{\circ}$) & $a$ & $\alpha$ & $\alpha_{AV}$\\
\hline
PS1 & 30 & 0.1 & 0.01 & 0.395\\
PS2 & 30 & 0.1 & 0.03 & 1.186\\
PS3 & 30 & 0.3 & 0.01 & 0.395\\
PS4 & 30 & 0.3 & 0.03 & 1.186\\
PS5 & 30 & 0.5 & 0.01 & 0.395\\
PS6 & 30 & 0.5 & 0.03 & 1.186\\
PS7 & 30 & 0.7 & 0.01 & 0.395\\
PS8 & 30 & 0.7 & 0.03 & 1.186\\
PS9 & 30 & 0.9 & 0.01 & 0.395\\
PS10 & 30 & 0.9 & 0.03 & 0.186\\
\hline
A1 & 0 & 0.9 & 0.01 & 0.395\\
A2 & 15 & 0.9 & 0.01 & 0.395\\
A3 & 30 & 0.9 & 0.01 & 0.395\\
A4 & 45 & 0.9 & 0.01 & 0.395\\
A5 & 60 & 0.9 & 0.01 & 0.395\\
A6 & 90 & 0.9 & 0.01 & 0.395\\
A7 & 120 & 0.9 & 0.01 & 0.395\\
A8 & 150 & 0.9 & 0.01 & 0.395\\
\hline
\label{table:simulations}
\end{tabular}
\end{center}
\caption{Simulation parameters, including the spin ($a$) and \citet{shakura_sunyaev} viscosity ($\alpha$) and artificial viscosity ($\alpha_{AV}$). Unless otherwise noted, the accretion discs also had $H/R=0.05$, an outer radius of $40R_{\rm in}$ and made use of $10^7$ particles.}
\end{table}

\subsection{Initial Conditions and Scope}
\label{subsection:scope}
Unless otherwise stated, the discs presented all made use of $10^7$ particles, and simulations with $ 10^6$ and $10^5$ were also conducted to check convergence. We note that each time the resolution is changed between simulations with otherwise the same parameters, the artificial viscosity is altered according to the scaling described in \citet{lodato_2010} so that the discs have the same $\alpha$ independent of which resolution is used. The locally isothermal sound speed in the disc was set to $c_{\rm s}(R) = c_{\rm s,in}(R/R_{\rm in})^{-q}$ and the surface density profile $\Sigma (R) = \Sigma_{\rm in}(R/R_{\rm in})^{-p}$, where $p=3/2$ and $q=3/4$ to give a constant $\alpha$ viscosity in the disc and uniform resolution \citep{lodato_pringle_2007}. Each disc was initially set up aligned to the black hole spin, with the particles arranged using a Monte Carlo placement method. Each particle was then rotated by the inclination angle and assigned a velocity according to the following expression derived from Equation~\ref{equation:vcrossh},
\begin{equation}
v_{\phi} = \frac{v_{\rm k}^4}{c^3}\left[\sqrt{a^2+\frac{R^3}{ R_{\rm g}^3}} - a \right] \cos (\theta),
\label{equation:spin_correction}
\end{equation}
where $v_{\rm k}$ is the Keplerian orbital velocity. The discs were therefore initially tilted to the black hole spin, but not warped. The results presented below have time shown in orbits at the inner edge and show the tilt as a function of radius only. This was found from the simulations using the method outlined in Section 3.2.6 of \citet{lodato_2010} where we used $N=300$ spherical shells.
For all of the simulations the inner radius was set as $R_{\rm in}=4R_{\rm g}$, in order to compare to the \citetalias{lubow2002evolution} 1D code. At small radii we note that the absence of GR limits the validity of our results, and indeed the need to carefully account for relativistic effects is one of our findings.

\begin{figure}
\includegraphics [width=\columnwidth] {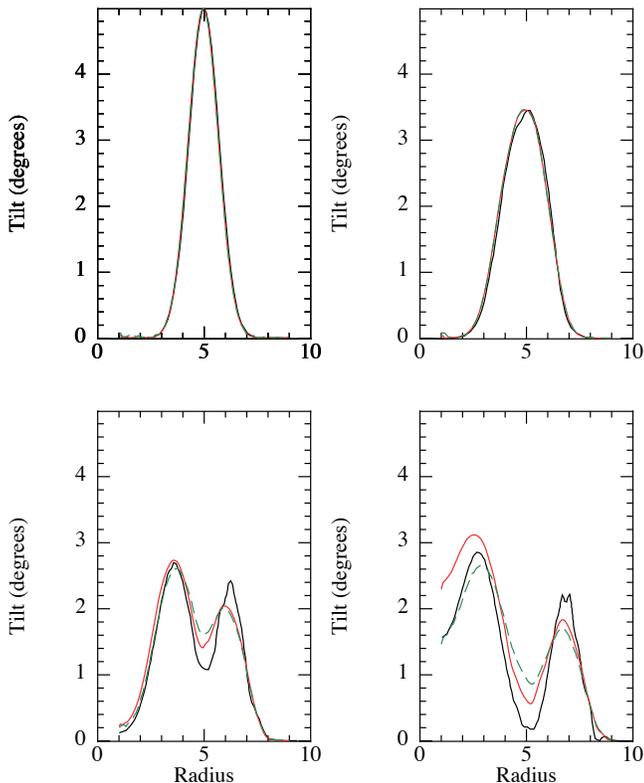}
\caption{Evolution of bending waves in a disc, not subject to Lense-Thirring precession. The green ($10^6$) and red ($10^7$) lines show the results from our 3D simulation. The black line shows the results from the 1D code of \protect\citet[Figure 1][]{Fragner:2010rt}, using the same initial parameters. The agreement between these two solutions confirms that SPH can be used to describe the evolution of warp propagation in the wavelike regime.}
\label{fig:FN_2010_match}
\end{figure}

\subsection{Test of wavelike warp propagation}
To confirm that we can correctly describe the propagation of warps in the wavelike regime, we use the test described by \citet{Fragner:2010rt}. They simulated a wavelike accretion disc with  a point mass potential and compared to a 1D calculation, finding agreement at the $\sim 10\%$ level. We choose to compare to this 1D solution instead of the solution from Equations~\ref{equation:wave1}~and~\ref{equation:wave2} because the linear solution from \citet{Fragner:2010rt} allows the surface density to evolve, as occurs in our 3D simulations.

We conduct a simulation using the same parameters as cited in their Figure 1, with $H/R=0.03$ and $\alpha=0.001$ and an initial disturbance of $5^{\circ}$. In this case we do not drive the evolution with Lense-Thirring precession, so that we can isolate the behaviour due to warp propagation only. Our results are shown in Figure~\ref{fig:FN_2010_match} using $10^6$ and $10^7$ particles. As the disc evolves, the disturbance splits into two waves travelling at half the sound speed (as predicted by \citealt{Papaloizou:1995pn}); one inward and the other outward. By the end of the simulation these have fully separated and are beginning to interact with the boundaries.

The SPH solution shows the same behaviour as the 1D solution, and increasing the resolution reduces the discrepancy. However, at late times the inner edge of the disc there is increasing disagreement, most likely due to differences in the inner boundary condition. This test confirms that \textsc{phantom} can be used to describe the propagation of warps in the wavelike regime.

\subsection{Test of Lense-Thirring precession}
We also perform a simple test of the Lense-Thirring precession. We simulate a disc consisting of test particles with no viscosity and zero sound speed (i.e. $\alpha=c_{\rm s}=0$) subject to Lense-Thirring precession. The initial velocities are set using Equation~\ref{equation:spin_correction} and the disc is inclined at $30^{\circ}$. We then calculate the precession in the disc as a function of the radius using the procedure outlined in Appendix~\ref{section:precession}.

Figure~\ref{fig:prec_example} shows the comparison between the precession measured from our disc and the predicted precession in the disc, given by $t_{\rm p} = R^3/(2a)$. We find agreement to within measurement uncertainties throughout the disc.

\begin{figure}
\includegraphics [width=\columnwidth] {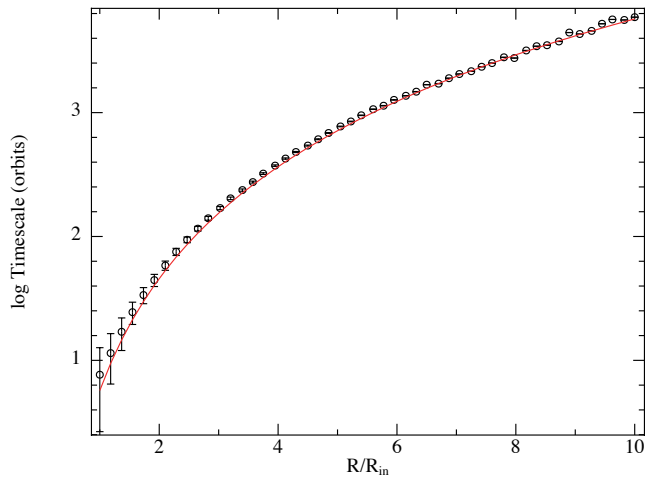}
\caption{Precession timescale measured from an inviscid and pressureless 3D disc as a function of radius (black circles), compared to the expected Lense-Thirring precession (red line). $R_{\rm in}=4R_{\rm g}$.}
\label{fig:prec_example}
\end{figure}

\begin{figure*}
\includegraphics [width=\columnwidth] {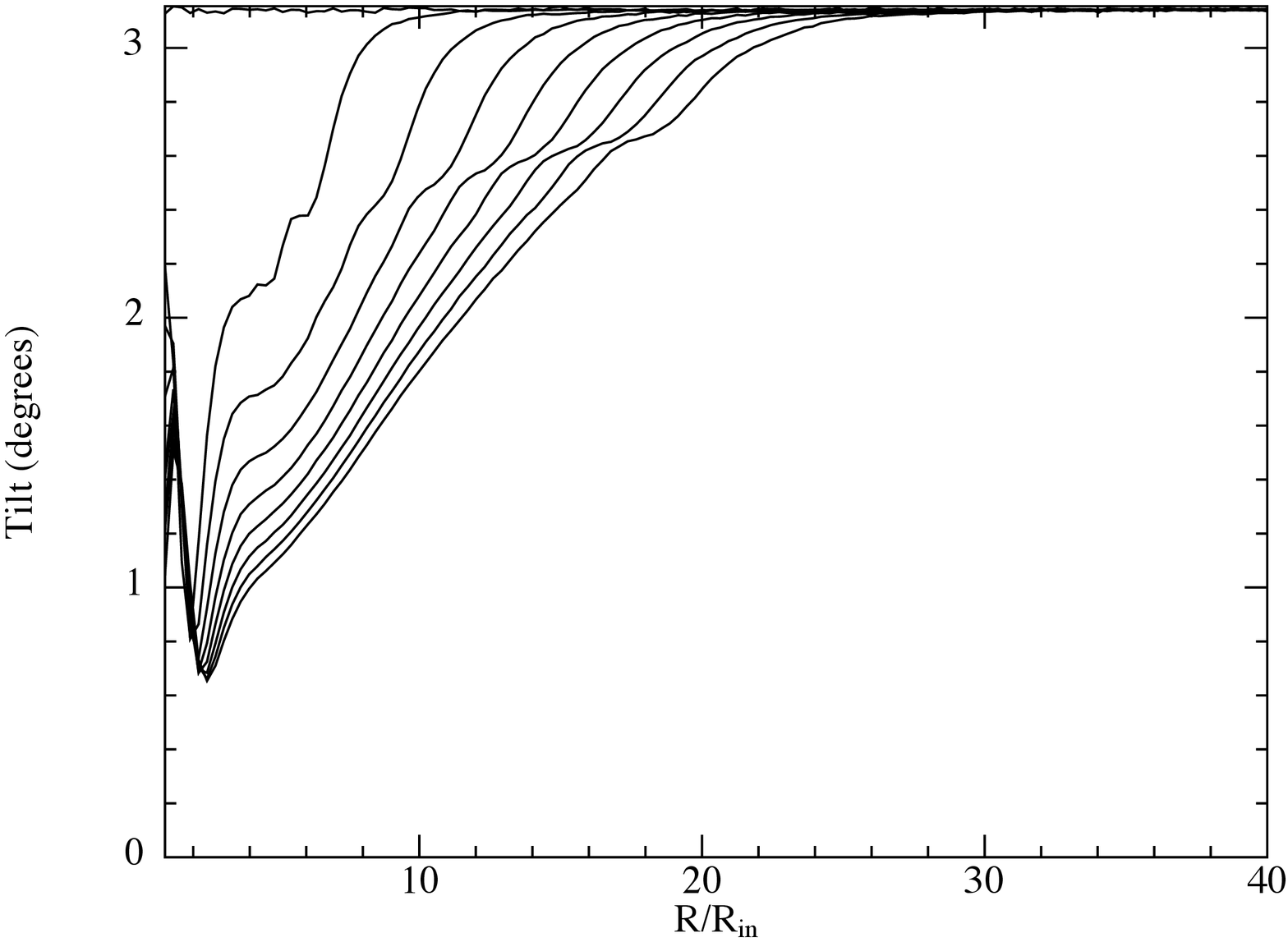}
\includegraphics [width=\columnwidth] {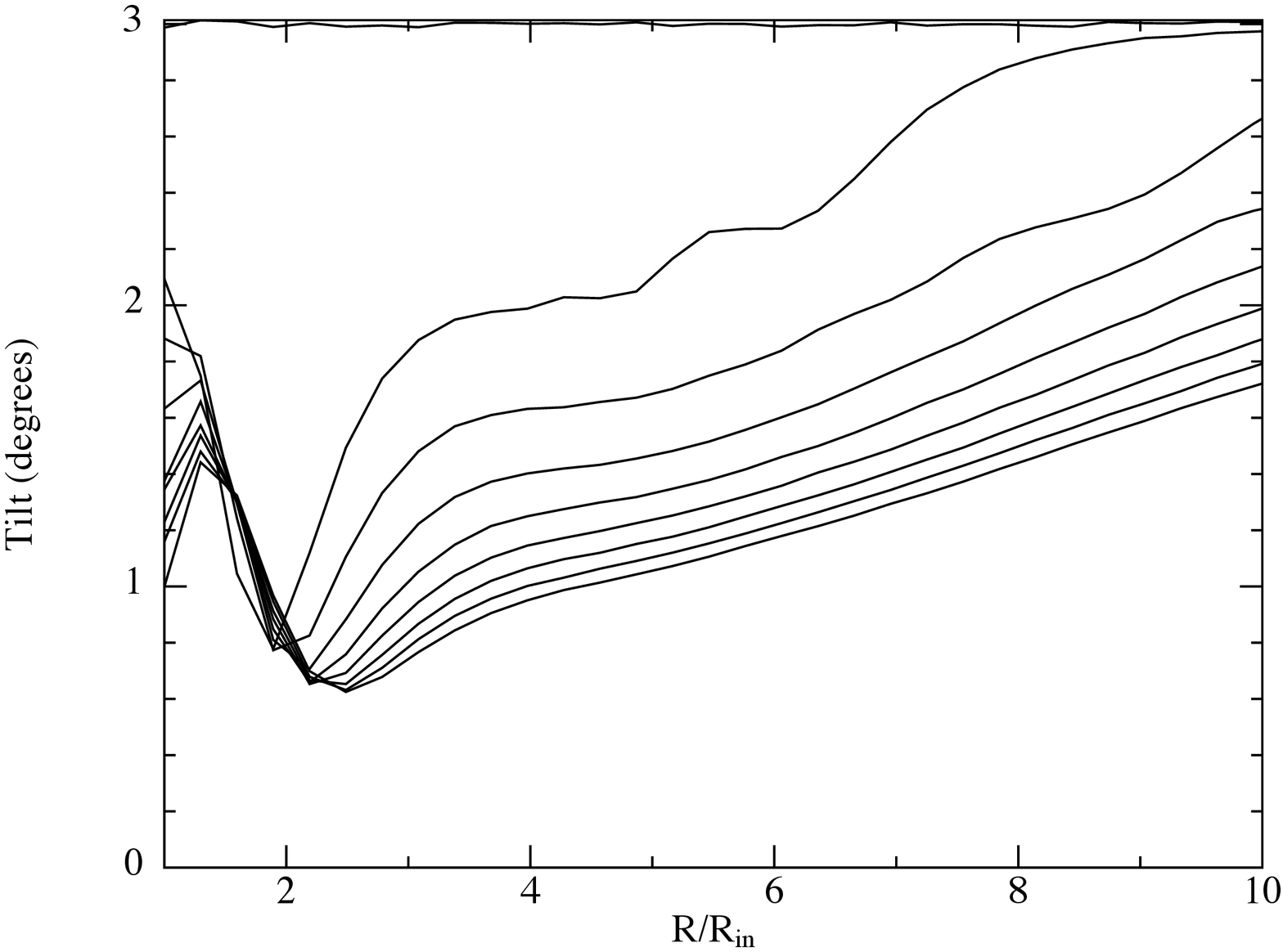}
\caption{Time evolution of the angle between the disc plane and the black hole spin as a function of radius in a 3D disc subject to Lense-Thirring precession, using similar parameters to \citet{lubow2002evolution} and the same times as in Figure~\ref{fig:LOP02_fig3}. The shape of this profile depends sensitively on the surface density at the inner edge and hence on resolution (see Figure~\ref{fig:surface_density}). The right panel shows a zoom-in of the inner disc.}
\label{fig:oscillation}
\end{figure*}

\begin{figure*}
\includegraphics [width=\textwidth] {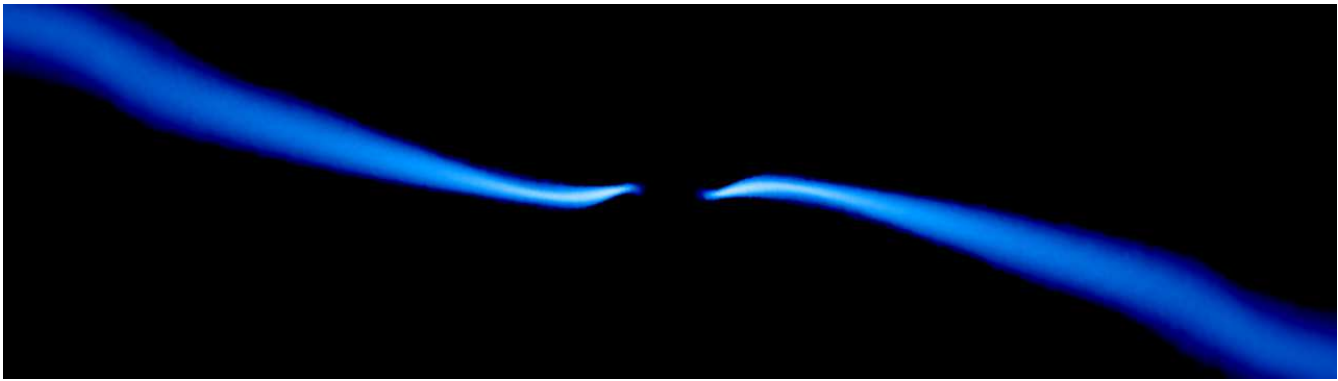}
\caption{Cross-section view of the steady-state tilt oscillation formed in a disc initially inclined at $15^{\circ}$ to the black hole spin (spin axis is vertical with respect to the page, i.e. along the $z$ axis). The colour scale shows density. Disc parameters are the same as in Figure~\ref{fig:oscillation}, but with a larger initial inclination.}
\label{fig:oscillation_cross_section}
\end{figure*}

\section{Results}
\label{section:results}
\subsection{Tilt Oscillations}
\label{subsection:tilt}
We first investigated whether or not the tilt oscillations predicted by \citet{lubow2002evolution} are physical using 3D simulations. The disc is initiated with a constant misalignment of $3^{\circ}$, within the linear regime required by Equations~\ref{equation:wave1}~and~\ref{equation:wave2}. We chose parameters for our simulation similar to that of \citet{lubow2002evolution}, with the exception of the surface density profile, the black hole spin and the disc thickness. Additionally, we made use of the Einstein potential outlined in Equation~\ref{equation:einstein_pot}, in order to give precession frequencies of the same sign (Equations~\ref{equation:eta_full}~and~\ref{equation:zeta_full}). We set $p=1.5$ and $q=0.75$ so that the disc is uniformly resolved, as discussed in Section~\ref{subsection:scope}. The disadvantage is that this results in lower amplitude oscillations in the 1D code. To combat this we encourage larger amplitude oscillations by increasing the spin to $a=0.9$ and decreasing the disc thickness to $H/R=0.05$. The evolution of the tilt as a function of radius is shown in Figure~\ref{fig:oscillation} from a simulation employing $10^7$ particles.

One of the main differences between this simulation and those conducted by \citet{nelson_pap_2000} is the angle of inclination. While our $3^{\circ}$ initial tilt was well in the linear regime required by the analytic description in Equations~\ref{equation:wave1}~and~\ref{equation:wave2}, the minimum inclination used by \citet{nelson_pap_2000} was $10^{\circ}$. We explore the effect of non-linear inclinations in this potential by misaligning the same disc at $15^{\circ}$. Figure~\ref{fig:oscillation_cross_section} shows a cross section of density in the inner disc from this calculation. The tilt profile after $\sim600$ orbits is qualitatively similar to Figure~\ref{fig:oscillation}, showing the same evolution. The quantitative tilt evolution is resolution-dependent, but nevertheless a non-zero tilt and oscillations were found at both medium ($10^6$ particles) and high ($10^7$ particles) resolution.

\begin{figure}
\includegraphics [width=\columnwidth] {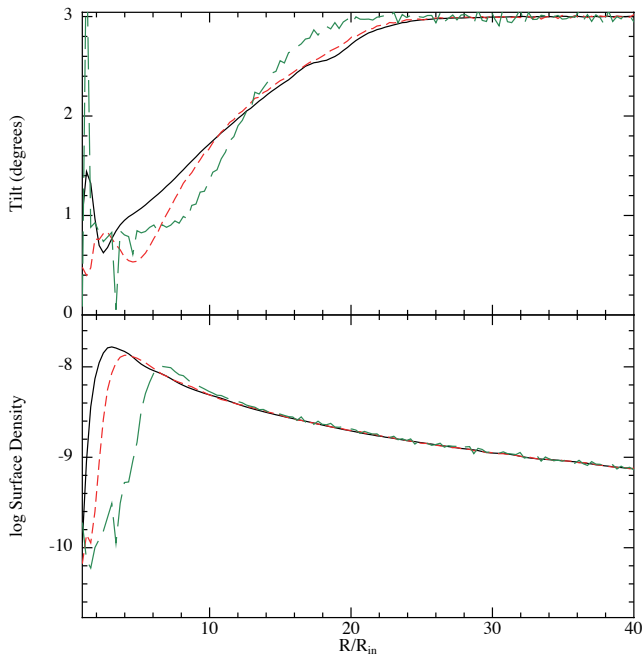}
\caption{Resolution study showing the inclination (tilt angle) and surface density as a function of radius at the final time of the disc shown in Figure~\ref{fig:oscillation}, using $10^5$ (long dashed green), $10^6$ (short dashed red) and $10^7$ (solid black) particles. Increasing the resolution better resolves the surface density profile at the inner edge, which strongly affects the final tilt profile found.}
\label{fig:surface_density}
\end{figure}

Figure~\ref{fig:surface_density} shows a resolution study of the tilt and surface density profiles using $10^5$, $10^6$ and $10^7$ particles. The main artefact of low resolution is that accretion occurs faster and as a result there is less mass at the inner edge in the lower resolution calculations. Comparison of the surface density profiles indicates that $\Sigma(R)$ is not fully converged near the inner edge, which has a dramatic effect on the tilt profiles. However, in all discs there is a non-zero tilt at the inner edge and in the $10^6$ and $10^7$ discs radial oscillations are observed. The wavelength of these oscillations is consistent with the criteria given by \citet{lubow2002evolution}. Even using the precession frequencies and surface density profiles in the 1D code that are appropriate to the 3D simulations (Equations~\ref{equation:eta_full}~and~\ref{equation:zeta_full}) still does not provide a close match with the 3D results. It is not clear if this discrepancy is due to non-linear fluid effects, e.g. as discussed by \citet{nelson_pap_2000}, or simply requires higher resolution calculations to obtain numerically converged results. However it is clear that with the appropriate potential and system parameters, the disc can display radial tilt oscillations as predicted by \citet{i_and_i_1997} and \citetalias{lubow2002evolution}.

\begin{figure}
\includegraphics [width=\columnwidth] {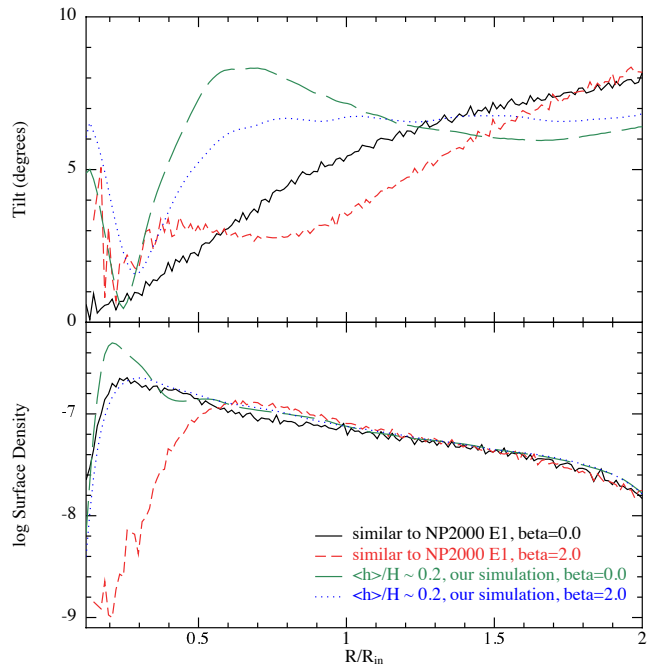}
\caption{The effect of bulk viscosity on the final tilt of the accretion disc. The solid, black line uses the same parameters as simulation E1 of \protect\citet{nelson_pap_2000}, with $5.2 \times 10^4$ particles and has no bulk viscosity. Bulk viscosity is used with both the red (short dashes, $5.2 \times 10^5$ particles) and green (long dashes, $5.2 \times 10^6$ particles) lines. At higher resolution and with bulk viscosity tilt oscillations are resolved, but the innermost parts of the disc remain unconverged. Here $R_{\rm g}=0.04$ and the disc has been evolved until the warp has reached the outer edge.}
\label{fig:matching_NP2000}
\end{figure}

Despite the resolution-dependence of our results, we were still able to observe tilt oscillations at resolutions used by \citet{nelson_pap_2000} so long as Einstein precession was accounted for. We further investigated whether this might be due to the differences in the artificial viscosity parameters used, as we set the Von Neumann-Richtmyer viscosity coefficient $\beta_{\rm AV}=2.0$ (see \citealt{Price:2012fj}) for all of our simulations whilst \citet{nelson_pap_2000} used $\beta_{\rm AV}=0$. A nonzero $\beta_{\rm AV}$ viscosity is required to prevent particle penetration \citep{monaghan_1989} and the absence of bulk viscosity is known to be problematic in disc simulations \citep{lodato_2010}. Thus with $\beta_{\rm AV}=0$, the simulations of \citet{nelson_pap_2000} might not have captured the wave interactions that create the tilt oscillations and the absence of bulk viscosity. To check this we conduct a low-resolution simulation equivalent to simulation E1 of \citet{nelson_pap_2000} and $\beta_{\rm AV} = 0$. Figure~\ref{fig:matching_NP2000} shows the results (black solid line), compared to an equivalent simulation with $\beta_{\rm AV} = 2$ (red dashed line) and also compared to a higher resolution simulations. At high resolution we find tilt oscillations regardless of the value of $\beta_{\rm AV}$ (green solid and blue dotted lines) but we find that using $\beta_{\rm AV} = 0$ can indeed erase the tilt oscillations at low resolution. The lower panel of Figure~\ref{fig:matching_NP2000} shows that this is not simply due to the effect on $\Sigma(R)$, since two of the calculations show very similar surface density profiles but rather different evolutions of the inner disc tilt.

\subsection{When does the disc break?}
\label{subsection:tearing}
\subsubsection{Bardeen-Petterson Alignment}
\label{subsection:a_and_alpha}
A second possible violation of the Bardeen-Petterson picture may be that the disc breaks instead of maintaining a smooth transition between an aligned inner disc and a misaligned outer disc. In order to investigate this, we simulate a range of discs at $30^{\circ}$ whilst varying $\alpha$ and $a$ according to the list PS1-10 in Table~\ref{table:simulations}. Here, for simplicity, we make use of a standard potential given by Equation~\ref{equation:pn_pot}, and hence do not expect any oscillatory behaviour.

\begin{figure*}
\begin{center}
\includegraphics [width=\textwidth] {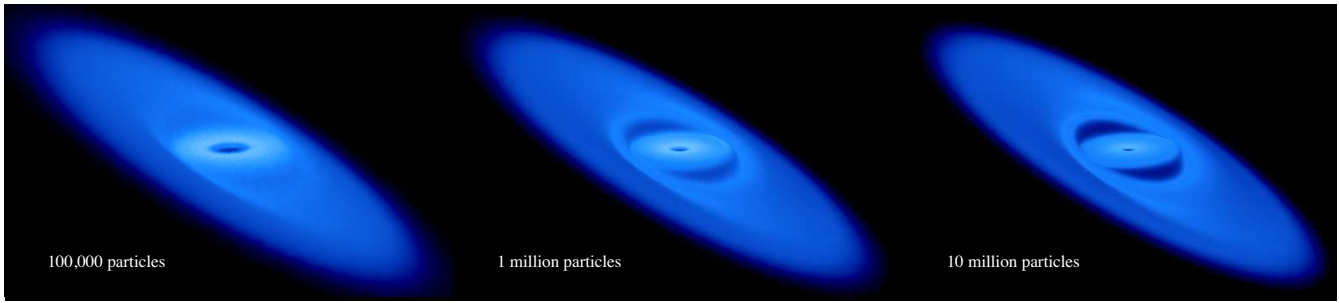}
\caption{Structure of the disc with $a=0.9$ and $\alpha=0.03$ at increasing numerical resolution (left to right). At low resolution the Bardeen-Petterson Effect is observed, similar to the results of \protect\citet{nelson_pap_2000}, but at high resolution the disc distinctly tears into two separate sections. The colour indicates density, with white being highest.}
\label{fig:res_study}
\end{center}
\end{figure*}

 Figure~\ref{fig:res_study} shows a 3D rendering of density in one such simulation with $a=0.9$ and $\alpha=0.03$ at three different resolutions (PS10). Except for the potential used, this disc has similar parameters to simulation E3 of \citet{nelson_pap_2000} which made use of $52,000$ particles across a larger radial extent than our simulations, representing a lower resolution than any of those shown in Figure~\ref{fig:res_study}. At our lowest resolution (left panel of Figure~\ref{fig:res_study}), we also observe the inner disc aligning and smoothly transitioning to an outer, misaligned disc (see their Figure 12). However, at higher resolutions this behaviour is no longer observed --- the disc instead breaks into two distinct sections, with the inner disc aligned with the black hole spin and the outer disc remaining misaligned.

\begin{figure}
\includegraphics [width=\columnwidth] {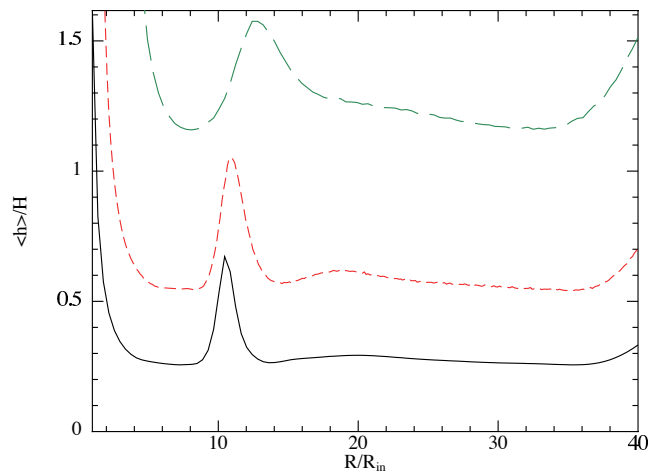}
\caption{Resolution length as a fraction of the disc scale height ($\langle h\rangle /H$) for the three resolutions ($10^5$ in green long dashes, $10^6$ in short red dashes and $10^7$ in solid black) shown in Figure~\ref{fig:res_study}. As disc breaking occurs on length scales smaller than the scale height of the disc, the lowest resolution simulations here cannot resolve breaking behaviour (as the resolution length is greater than the scale height throughout the disc). The two higher resolution simulations are able to resolve this behaviour.}
\label{fig:smoothing_length}
\end{figure}

\begin{figure*}
\begin{center}
\includegraphics [width=\textwidth] {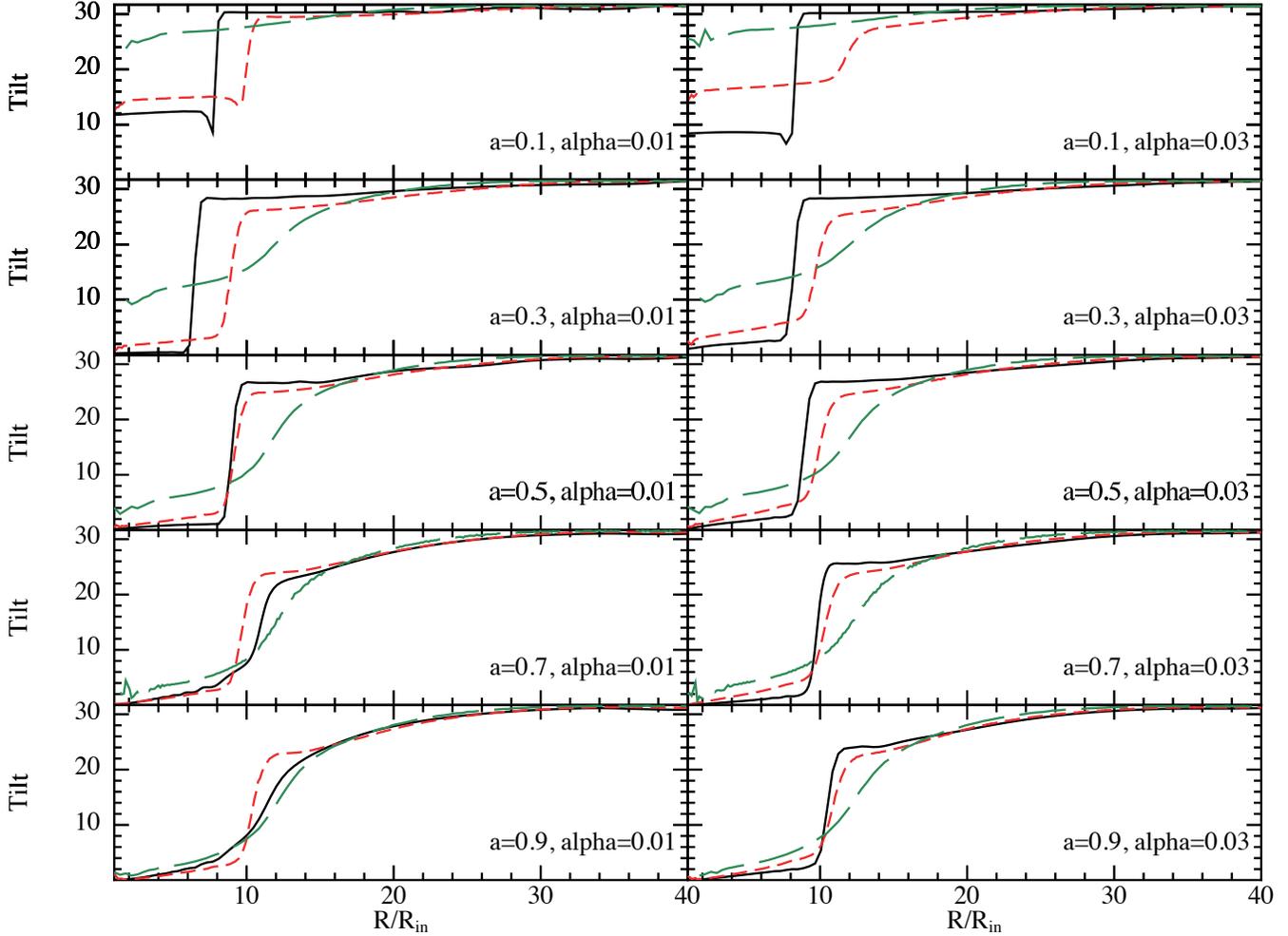}
\caption{Disc tearing for different spin and viscosity combinations, with the same initial tilt of $30^{\circ}$ and $10^5$ particles shown in green (wide dashes), $ 10^6$ particles shown in red (dashes) and $ 10^7$ particles shown in black (solid). At the lowest resolution we observe the Bardeen-Petterson effect complete with a smooth transition for most discs. Increasing the resolution results in disc tearing (Bardeen-Petterson alignment) independent of our choice of viscosity or spin. As we do not include the affect of Einstein precession, we do not observe radial tilt oscillations in these discs.}
\label{fig:parameter_sweep_tilt}
\end{center}
\end{figure*}

 Figure~\ref{fig:smoothing_length} shows that resolving disc breaking is mainly a question of resolving the disc scale height. For the lowest resolution simulations the resolution length is greater than the scale height of the disc and hence disc breaking (on a length scale smaller than $H$) cannot be resolved and a smooth transition is observed. By contrast, the two higher resolutions are able to resolve disc breaking.
 Figure~\ref{fig:parameter_sweep_tilt} shows the same resolution study performed in Figure~\ref{fig:res_study} for all of our discs at $30^{\circ}$, where the green line shows simulations that made use of $10^5$ particles, red shows $ 10^6$ and black shows $10^7$ particles. The discs are shown after $1500$ orbits at the inner edge, allowing the warp to propagate all the way to the outer radius. Increasing the spin of the black hole increases the rate at which the innermost part of the disc aligns.

Across all of the parameters chosen here, increasing the resolution changes the behaviour from the smooth tilt profile observed by \citet{nelson_pap_2000} to a steepening of the tilt profile and ultimately a disc that is broken into distinct sections. The higher resolution results show an aligned inner edge, a misaligned outer edge and a sharp tilt profile connecting these, representing a break in the disc.  The discs simulated with lower spins appear to steepen and tear faster than those with higher spin as the break occurs further out (and hence a longer precession time). This is observed most clearly between the low viscosity, high spin cases. At $a=0.5$, the tilt steepened and the disc tore before the end of the simulation. For the disc with $a=0.7$, the tilt began steepening near the end of the simulation but was not able to separate, whilst at $a=0.9$ the disc has not yet begun steepening. We have confirmed that this is the case by extending the high resolution simulations of the $a=0.9$ case, and indeed observed steepening to occur at later times.

As with the previous simulations, Figure~\ref{fig:parameter_sweep_tilt} demonstrates that the simulations are not fully converged, especially when considering the low spin cases ($a<0.5$). For these discs, the discrepancy in the inner tilt is again due to the mass accreted at the inner edge of the disc. At low resolutions the inner part of the disc is accreted faster, resulting in less mass near the inner edge. The same Lense-Thirring torque then acts on less mass, and is thus not able to align the disc to the same extent. At increasing spins this effect is observed less, as the higher spin provides a larger torque and so even the lowest resolution discs are able to align. In the discs with $a<0.5$, increasing the resolution leads to a more distinct tear in the disc suggesting that our results are consistent. Hence we can be confident that these discs do tear, and present an upper limit on the radius at which this occurs. As the tearing occurs outside of the radius where oscillations were found in Section~\ref{subsection:tilt}, using similar parameters, this behaviour should not be affected by our choice of potential.

\subsubsection{Disc Tearing}
\begin{figure*}
\begin{center}
\includegraphics [width=0.7\textwidth] {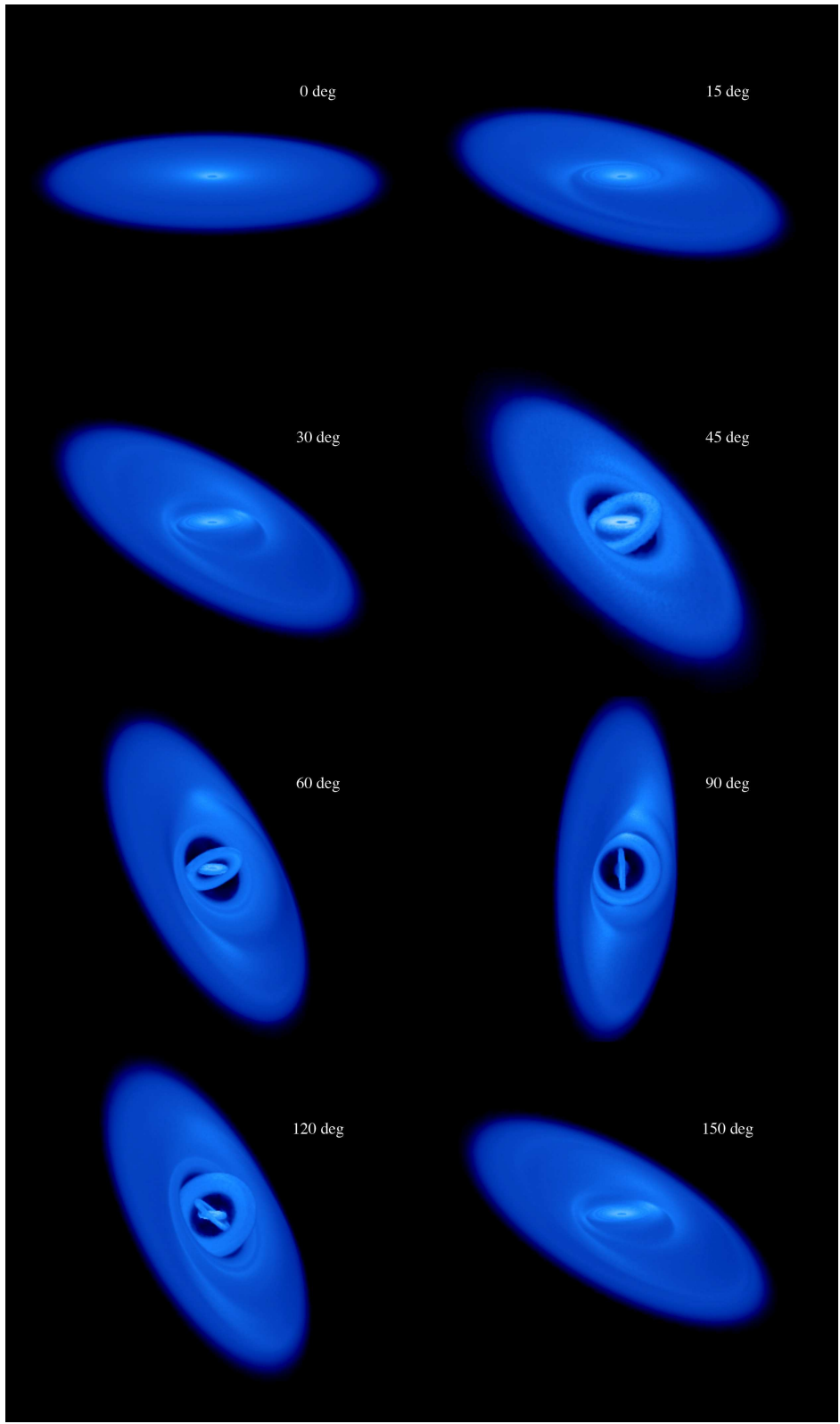}
\caption{3D renderings of discs that were initially misaligned with the black hole spin at various angles, with each simulation using $10^7$ particles and shown after $\sim1500$ orbits. The inability of the discs inclined by more than $\theta \gtrsim 45^{\circ}$ to communicate the Lense-Thirring precession causes the formation of discrete rings which `tear' and precess effectively independently before undergoing direct cancellation of angular momentum and rapid accretion. The black hole spin in each of these images is vertical with respect to the page (i.e. along the $z$ axis). The same density scale is used as in Figure~\ref{fig:res_study}.}
\label{fig:tilt_angle}
\end{center}
\end{figure*}

To investigate the dependence of disc tearing on the misalignment between the disc and the black hole spin in the wavelike regime we simulated a suite of discs at different inclinations. We again make use of the traditional post-Newtonian approximation given by Equation~\ref{equation:pot}. We held $\alpha=0.01$ and $a=0.9$ constant and varied the inclination of the disc between $0^{\circ}$ (aligned) and $150^{\circ}$, noted in Table~\ref{table:simulations} with A1-8. Figure~\ref{fig:tilt_angle} shows these simulations after more than $1500$ orbits measured at the inner edge. Each disc was initially tilted but not warped. As the simulation progressed, a warp evolved in response to the Lense-Thirring torque and in the higher inclination cases resulted in the disc breaking.

At $15^{\circ}$ (top right of Figure~\ref{fig:tilt_angle}) the disc was observed to smoothly align to the spin of the black hole. At the end of the simulation, the tilt of the disc was consistent with the Bardeen-Petterson effect and is similar to results seen in previous simulations at $10^{\circ}$ by \citet{nelson_pap_2000}. Extending the lower resolution version of this simulation (with $10^6$ particles) for twice as long shows that the disc continues to align with the black hole spin, implying that the steady state for this disc is full alignment. Inclining the disc at $30^{\circ}$ also did not yet result in disc tearing, however this is because for this particular choice of viscosity and spin this simulation has not been run long enough (see Section~\ref{subsection:a_and_alpha})

For discs at higher inclinations ($\gtrsim 45^{\circ}$; second, third and fourth rows of Figure~\ref{fig:tilt_angle}), the inner section of the disc was found to align within $50$ orbits and a smooth transition was formed between this and the outer region of the disc. This transition then steepened until the disc broke into two sections that were connected by precessing rings of material. Multiple rings of material were torn off from the outer, misaligned disc and each was observed to precess effectively independently.  Towards the end of the simulations, up to two rings were precessing at the same time (for example, $120^{\circ}$ disc of Figure~\ref{fig:tilt_angle}) and were present for up to $\sim400$ orbits. Eventually each of these rings settled with and increased the inner, aligned region of the disc. The disc inclined at $90^{\circ}$ (right hand panel in third row of Figure~\ref{fig:tilt_angle}) also developed precessing rings of material that were accreted. However, for this inclination no inner aligned disc was observed and the rings of material were accreted directly onto the black hole.

\begin{figure}
\includegraphics [width=\columnwidth] {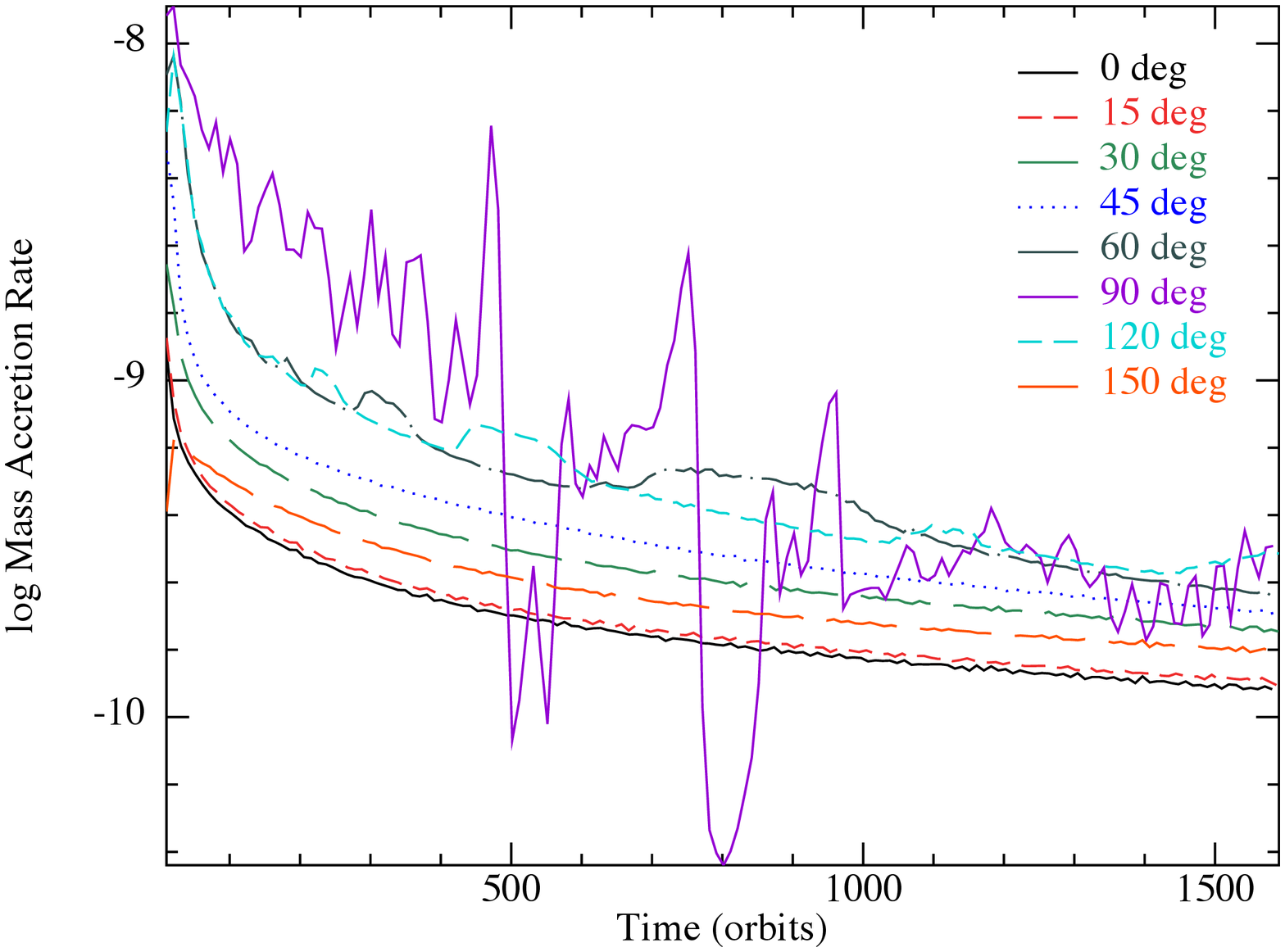}
\caption{Instantaneous mass accretion of the $0^{\circ}$, $15^{\circ}$, $30^{\circ}$, $60^{\circ}$, $90^{\circ}$, $120^{\circ}$ and $150^{\circ}$ discs run with $10^6$ particles and time measured in orbits.  The bin width is 10 times the orbit timescale and a logarithmic scale is used for convenience. In line with previous results, inclining the disc to the black hole results in mass accreting faster by almost an order of magnitude.}
\label{fig:mass_accretion}
\end{figure}

Figure~\ref{fig:mass_accretion} shows the instantaneous mass accretion rate by the discs at different angles. It can be seen that inclining the disc to the spin of the black hole increases the rate of accretion by more than an order of magnitude when compared to an aligned disc, similar to previous findings \citep{nixon_et_al_2012}. The discs that form an inner aligned disc and precessing rings have even higher accretion rates, as the inner disc is continually fed by the rings as they align. When taken in context with the results in the diffusive regime \citep{nixon_et_al_2012}, this implies that regardless of whether the disc is thin or thick, mass accretion is faster when the disc is inclined.

Disc tearing has also been observed in the wavelike disc regime for circumbinary discs inclined at high angles \citep{facchini_2013}. In a simulation of a circumbinary disc inclined at $60^{\circ}$, their disc separates into two sections and the inner one precessed effectively independently of the outer disc. As their disc is thicker than ours ($H/R=0.1$) and has a higher viscosity ($\alpha=0.05$), we would anticipate that a strong external torque would be required to tear the disc, and we observe their disc does tear at a smaller radius than any of ours.

\subsubsection{Location of tearing radius}
The disc is expected to tear when the Lense-Thirring torque is larger than the internal communication in the disc. If the internal communication in the disc is governed by viscosity, the torques given in Section~\ref{subsection:torques} can be used to estimate the upper breaking radius given in Equation~\ref{equation:breaking_radius}. However in our simulations the disc internal dynamics are dominated by pressure rather than viscosity, hence Equation~\ref{equation:other_breaking_radius} may be more appropriate.

\begin{figure}
\includegraphics [width=\columnwidth] {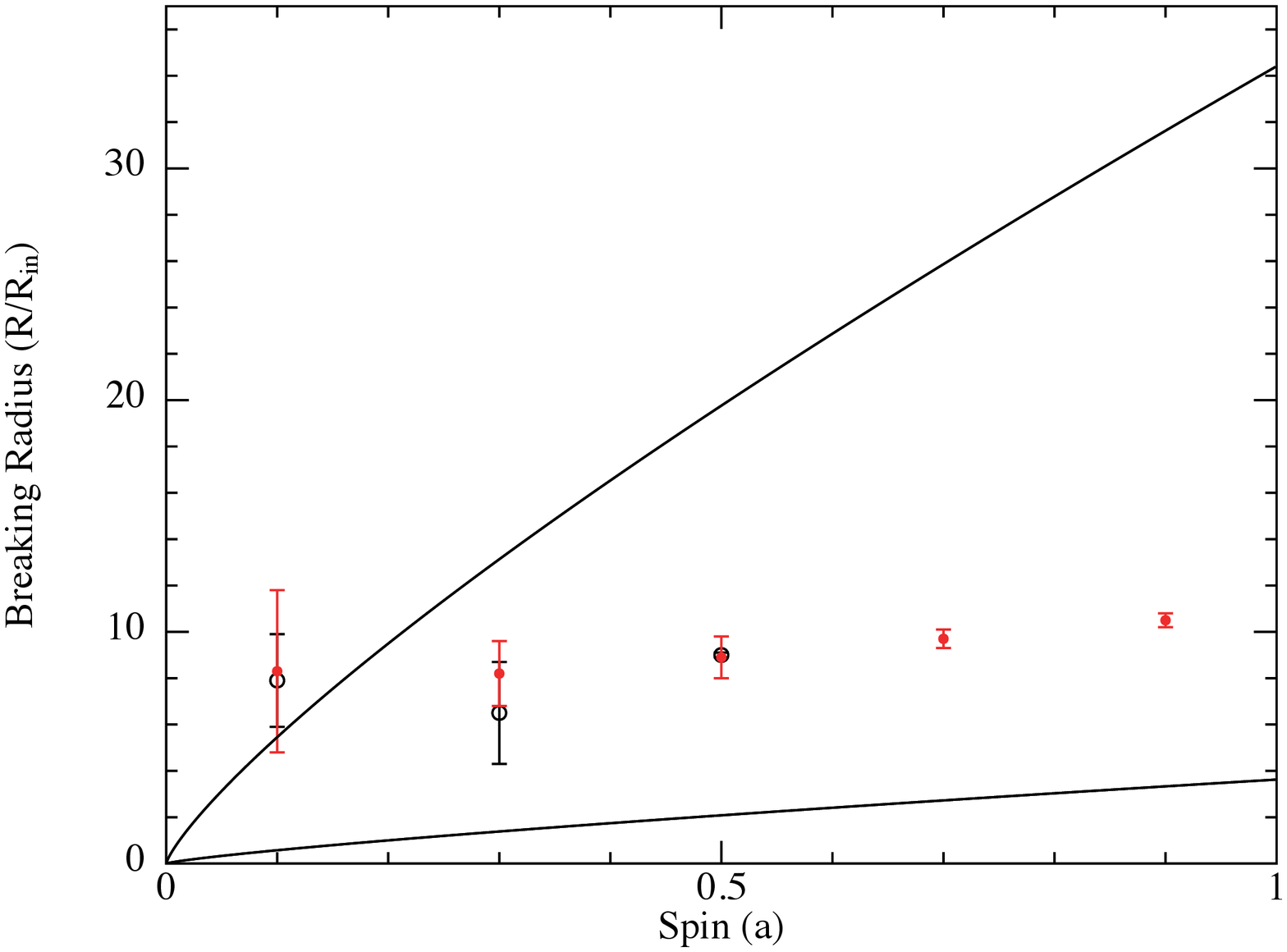}
\caption{Comparison of the breaking radius measured from the discs inclined at $30^{\circ}$ with our prediction of $R_{\rm break}$ (upper) found by considering the torques in the disc (Equation~\ref{equation:breaking_radius}) and $R_{\rm break,t}$ (lower) by comparing the sound crossing and precession timescales (Equation~\ref{equation:other_breaking_radius}).}
\label{fig:r_break}
\end{figure}

 As the Lense-Thirring torque has a radial dependence, it is largest in the inner most parts of the disc and it is reasonable that these discs will break at a radius smaller than predicted by Equation~\ref{equation:breaking_radius}. Figure~\ref{fig:r_break} shows a comparison of the estimated break radius for the simulations inclined at $30^{\circ}$ compared to the prediction from Equation~\ref{equation:breaking_radius} (upper line; assuming that $\alpha=0.02$, the average for our simulations) and from Equation~\ref{equation:other_breaking_radius} (lower line). We find that the disc does break at radii lower than our prediction from the viscous torques alone, and that the breaking radius is intermediate between the predictions from Equations~\ref{equation:breaking_radius} and \ref{equation:other_breaking_radius}, indicating that the torques in our discs lie between these two extremes. The increasing uncertainties at low spin correspond to the decreasing convergence of our simulations due to mass accretion at the inner edge, seen in Figure~\ref{fig:parameter_sweep_tilt}.

The discrepancy between the predicted and the observed breaking radius appears to occur at all inclinations. Using Equation~\ref{equation:breaking_radius}, the breaking radius for the $60^{\circ}$ disc is found to be $R_{\rm break}\sim 41 R_{\rm in}$ which is greater than $R_{\rm out}$. However this disc is observed to break (at $R \lesssim 18 R_{\rm in}$), in line with the results of Figure~\ref{fig:r_break}. If we now consider the $15^{\circ}$ disc, it is predicted to break at $R_{\rm break}\sim 18 R_{\rm in}$ but from the simulation we do not observe tearing. This could occur if the actual tearing radius is less than $R_{\rm in}$, consistent with the previous results.

\subsubsection{Width of the rings}
\begin{figure}
\includegraphics [width=\columnwidth] {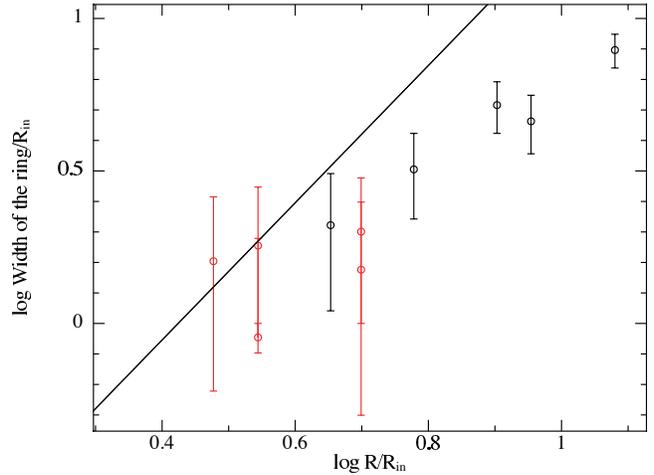}
\caption{The solid line shows the expected width of each ring of gas torn off in our simulations, calculated by comparing the precession timescale to the distance that the wave can travel (Equation~\ref{equation:ring_width}). The circles show the measured ring widths from the simulations, where black circles indicate short lived rings and red circles rings that are stable for more than $\sim20$ orbits.}
\label{fig:ring_width}
\end{figure}

The rings that are torn off during the simulations appear to be much wider than those found in the diffusive regime \citep{nixon_et_al_2012}, some up to $\Delta R/H(R) \sim 25$ (where $\Delta R$ represents the ring width). It is possible for rings to form when the disc is able to break and differential precession is present, such as when the disc is subjected to Lense-Thirring precession. We therefore expect the width of the ring to be determined by a relative comparison between the sound crossing and precessional timescales in the disc. We can approximate this by letting $\Delta R$ be the distance that a wave can travel in a precession time such that 
\begin{equation}
\int\frac{2}{c_{\rm s}} dR \propto t_{\rm p},
\end{equation}
across the ring. If we assume that the inner edge of the ring is at $R_{\rm in}=R-\Delta R/2$, the outer edge at $R_{\rm out}=R+\Delta R/2$ and use the expression for the sound speed, we get
\begin{equation}
R^3 \propto \frac{a}{(q+1)} \left[1 - \left(\frac{R_{\rm in}}{R_{\rm out}}\right)^{q+1}\right] \frac{R_{\rm out}}{c_{\rm s}(R_{\rm out})}, \label{equation:ring_width}
\end{equation}
where $R$ is the radius that a ring of thickness $\Delta R$ occurs at. Figure~\ref{fig:ring_width} compares the width of the rings measured from the simulations to this prediction. Although there are large uncertainties in the measurements the general trend of increasing ring width with $R$ is reproduced.

\label{subsubsection:tearing_radius}

\begin{figure*}
\includegraphics [width=\textwidth] {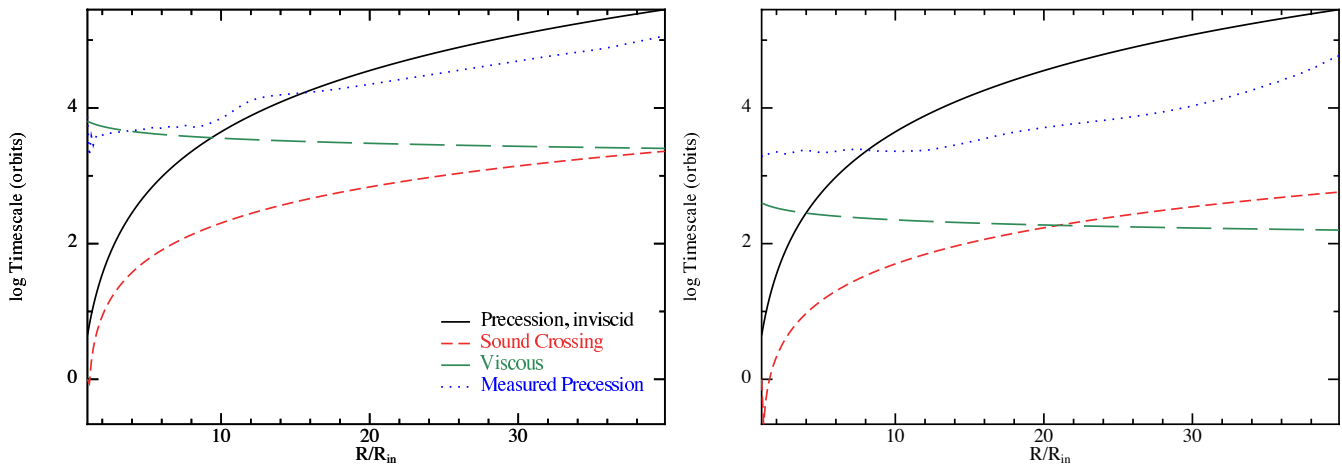}
\caption{The timescales in our PS9 simulation (left) and a disc that is four times thicker (right). The precession timescale does not change much between the thin and thick discs, however the sound crossing and viscous timescales decrease as the disc becomes thicker. The decrease in viscous time means that the thicker disc is able to accrete misaligned material, preventing the development of a steady state.}
\label{fig:time_scales}
\end{figure*}

\subsection{Can the disc accrete misaligned?}
\label{subsection:mass}
Previous simulations of tilted accretion discs in the wavelike regime have not identified disc tearing when the disc is subjected to Lense-Thirring precession. The results of these thicker discs have found that the disc warps in the inner region, with a non-zero tilt at the inner edge, and then precesses as a solid body \citep{fragile_anninos_2005,Fragile:2007uq}. To examine this behaviour, we conduct a single simulation of a thick disc. We use the same parameters as PS9, but with an aspect ratio four times the initial value, such that $H/R=0.2$ at the inner edge. This disc is similar to the simulation of \citet{Fragile:2007uq}, except that it has twice the initial tilt (and does not include magnetic fields).

The timescales for this disc are shown in the right of Figure~\ref{fig:time_scales}. Comparison with the timescales from PS9 (left) shows that although the precession timescale has not changed appreciably, the viscous and sound crossing timescales have decreased substantially. In the outer half of the disc we note that the viscous timescale is the shortest, allowing the material located there to be accreted to the inner regions faster than it can align. This leads to material being accreted before it can align with the spin of the black hole, causing a non-zero tilt at the inner edge of the disc. A comparison between the mass accretion of this disc and our thinner PS9 simulation shows that there is more mass accreted by the thicker disc.

\begin{figure}
\includegraphics [width=\columnwidth] {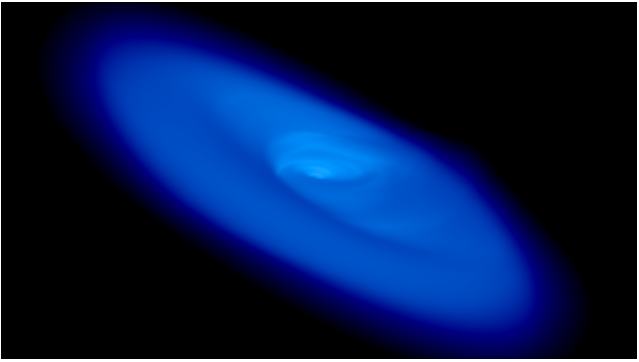}
\caption{Our thick disc simulation, similar to that of \protect\citet{Fragile:2007uq} except that it is initialised at $30^{\circ}$ and run for ten times longer. This disc is not observed to tear, as expected, but warping is observed in the inner regions and higher mass accretion than our thin disc. This figure is shown with the same density scale as Figures~\ref{fig:res_study}~and~\ref{fig:tilt_angle}.}
\label{fig:matching_FBAS07_SPH}
\end{figure}

\begin{figure}
\includegraphics [width=\columnwidth] {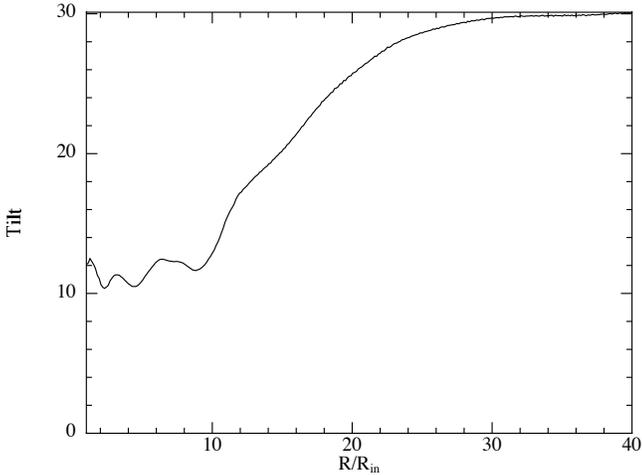}
\caption{The final tilt profile of our thick disc simulation. Misaligned accretion occurs at the inner edge, causing a non-zero tilt and preventing a steady state from being formed. The inner edge features are not steady like the results of Figure~\ref{fig:oscillation_cross_section}.}
\label{fig:matching_FBAS07}
\end{figure}

Simulating to approximately the same time as quoted by \citet{Fragile:2007uq}, we observe the thick disc to warp in the inner regions but not to tear. At this time in our thin disc simulations we also do not observe tearing, so we continue the simulation until approximately $200$ orbits according to the time units specified by \citet{Fragile:2007uq} ($10$ times longer than their lower resolution simulation). The results at this time are shown in Figure~\ref{fig:matching_FBAS07_SPH}~and~\ref{fig:matching_FBAS07}. We do not observe the large increase in the disc tilt at the inner edge that was found by \citet{Fragile:2007uq} (see their Figure 12), however in their paper this is attributed to plunging streams which we also do not observe. Presumably this is due to our use of the post-Newtonian approximation in Equation~\ref{equation:pot} and consistent with the discussion in Section~\ref{subsubsection:freq}.

As the disc is four times thicker than our simulation PS9, $\nu$ increases by a factor of 16 (even though $\alpha$ does not change). This increases the internal torque in the disc by the same factor (see Equation~\ref{equation:internal_torque}), but the external torque applied is the same as for our disc. This should make it much harder to tear the disc, and when we calculate the breaking radius using Equation~\ref{equation:breaking_radius} we find that it would be $R/R_{\rm in} \sim 3$, inside the region where misaligned accretion is occurring. Indeed, from our results in Section~\ref{subsubsection:tearing_radius}, we would not expect this disc to tear at all.

%

\section{Discussion}
\label{section:disc}
Despite using up to $10^7$ particles, the simulations that have been presented are not yet converged. As shown in Figure~\ref{fig:res_study} and comparison of our results with \citet{nelson_pap_2000}, increasing the resolution strongly affects the behaviour the disc displays. However, features like disc tearing and radial oscillations are present in both the medium and high resolution simulations and so we can draw conclusions about the qualitative behaviour. Additionally, whilst increasing the resolution decreases the breaking radius, it does so by a smaller amount each time, so we are confident that the measured tearing radii for our non-linear simulations is an upper limit and that our results are close to being converged.

 The main point of discussion is why our results differ to those found by \citet{nelson_pap_2000}. The two main factors are the numerical resolution and the viscosity parameter $\beta_{AV}$. Simulations we performed at comparable resolution to \citet{nelson_pap_2000} showed similar behaviour --- namely a smooth transition between the aligned and misaligned regions. However, when we increased the resolution we found that the behaviour changes and these discs tear into two disconnected sections (the main criterion being to adequately resolve the disc scale height). This implies that low resolution prevented \citet{nelson_pap_2000} from observing disc tearing. However, we also showed that the inclusion of a $\beta$ viscosity, even at low resolution, recovers steady-state oscillations in the tilt of the disc midplane with respect to the black hole spin axis similar to those predicted by the linear theory of \citet{i_and_i_1997} and \citetalias{lubow2002evolution}. This is in contrast to the findings in \citet{nelson_pap_2000}, where it was suggested that the tilt oscillations were short wavelength features which could be damped out by non-linear effects. As shown by our $15^{\circ}$ simulation and in agreement with \citetalias{lubow2002evolution}, we found that the wavelength of the radial oscillation is of the order of the radius and is not damped out by such effects.


The tilt oscillations that were found at linear inclinations do not match to the description of the 1D code by \citetalias{lubow2002evolution}, and increasing the resolution does not reduce the discrepancy. The difference is likely due to the 1D code assuming that the viscous timescale is negligibly large. In Section~\ref{subsection:mass} it is found that the mass accretion is not necessarily negligible, as for discs with a larger aspect ratio we found it is possible for the material to accrete to the inner regions of the disc faster than it is able to align. This causes the disc to accrete misaligned material, which prevents a steady state from being formed and confirms that it is not possible to produce a tilt profile such as that described by the Bardeen-Petterson effect if the viscous time is too short (as predicted by \citealt{Lodato:2006fk}). Recently the thinnest discs in relativistic simulations have been completed by \citet{Morales-Teixeira:2014lr}, with $H/R=0.08$. Their retrograde simulation showed partial alignment at the inner edge, but their prograde simulations displayed an inner edge tilt that was greater than the initial condition. It is also noted that the strength of the tilt oscillations depends on the disc thickness, and so thick discs (and tori) would display weak oscillations.

Perhaps the main caveat of our simulations is that we use an $\alpha$ viscosity to model the discs. Whilst a comparison between a purely hydrodynamical disc (with no explicit viscosity) and one where the viscosity is controlled by the MRI has shown that the behaviour of the disc is largely controlled by the hydrodynamic evolution \citep{Sorathia:2013fk}, for a complete picture of the disc evolution we should include magnetic fields to self-consistently generate a turbulent viscosity through the MRI. However it is not yet clear how the MRI will respond in the presence of a warp, especially at large angles. Additionally, we assume that our discs are vertically isothermal. Heating of the disc due to warping may further complicate this picture \citep{Ogilvie:2003yq}.

For tilt oscillations and efficient wave transport to occur we require $H/R > \alpha$ \citep{Papaloizou:1995pn,i_and_i_1997}. Black hole accretion discs are often expected to be geometrically thin and have $\alpha \sim 0.1$ \citep{king_et_al_2007}. However, for discs which are accreting either at very sub-Eddington ($\lesssim 0.1 L_{\rm Edd}$) or near-Eddington ($\gtrsim L_{\rm Edd}$) rates, the disc may become geometrically thick \citep{Narayan:2005lr}. So the simulations presented here may be most relevant to the low luminosity state of X-ray binaries where the disc can be thick and $\alpha$ may be significantly smaller than its usual outburst value \citep{Smak:1984fk,Meyer:1984qy}. They are also relevant to AGN accreting at rates greater than Eddington and to the discs formed in tidal disruption events where the initial star orbit can be highly misaligned and the disrupted material infalling at super-Eddington rates.



\section{Conclusion}
\label{section:conclusion}
In this work we have re-examined the Bardeen-Petterson effect in 3D using hydrodynamical simulations of accretion discs subject to Lense-Thirring precession, in the regime where warps propagate in a wavelike manner ($\alpha \lesssim H/R$). Our detailed conclusions are as follows:
\begin{enumerate}
\item The Bardeen-Petterson picture of an aligned inner disc smoothly connected to a misaligned outer disc occurs only at low inclinations and only when Einstein precession is not accounted for. Using high resolution calculations, we find both steady state oscillations in the disc tilt (when Einstein precession is included) and that discs break when they are relatively thin and highly misaligned to the black hole spin.
\item We recover steady tilt oscillations for the first time in a 3D hydrodynamics code, as predicted by \citetalias{lubow2002evolution}. However, as the 1D code developed by \citetalias{lubow2002evolution} assumes that mass accretion is negligible, discrepancies remain between the predicted tilt profile and our 3D results.
\item Tilt oscillations are also present at higher inclinations ($15^{\circ}$), showing that non-linear effects do not necessarily damp this behaviour.
\item Disc `tearing' or `breaking', rather than a smooth transition between spin-aligned and spin-misaligned parts of the disc, appears to be an inevitable outcome for accretion discs inclined to the black hole spin by more than a few degrees. This occurs regardless of whether the propagation of bending waves is governed by pressure forces or viscous stresses.
\item Tearing of the disc leads to rings that precess effectively independently. As in the diffusive regime, this can lead to direct cancellation of angular momentum and hence faster accretion. The main difference in the wavelike regime is that the rings are wider, with the width determined by the ratio of precession to sound crossing time rather than the disc scale-height.
\item The Bardeen-Petterson effect cannot occur in discs where the viscous time is comparable to the alignment time. In this case the disc material is accreted misaligned. Hence it is possible to have discs that are misaligned with respect to the black hole spin even in the absence of tilt oscillations, but this can only occur at high $\dot{M}$ (i.e. for thick discs).
\item Mass accretion rates can be enhanced by an order of magnitude or more when the disc is inclined with respect to the black hole spin. This occurs regardless of whether the disc is thick or thin.
\end{enumerate}

\section*{Acknowledgments} 
We thank the anonymous referee and Gordon Ogilvie for valuable comments that improved the manuscript. We thank Jim Pringle for supplying the original 1D code used to compute the \citetalias{lubow2002evolution} solution. We also thank Matthew Bate for useful discussions. RN is supported by an Australian Postgraduate Award. DJP is supported by a Future Fellowship (FT130100034) from the Australian Research Council. CN thanks NASA for support through the Einstein Fellowship Programme, grant PF2-130098. This work was performed on the gSTAR national facility at Swinburne University of Technology. gSTAR is funded by Swinburne and the Australian Government's Education Investment Fund. We used \textsc{splash} \citep{Price:2007kx} for the plots and visualisations.

\appendix
\section{Measuring the disc precession}
\label{section:precession}
\begin{figure}
\includegraphics [width=\columnwidth] {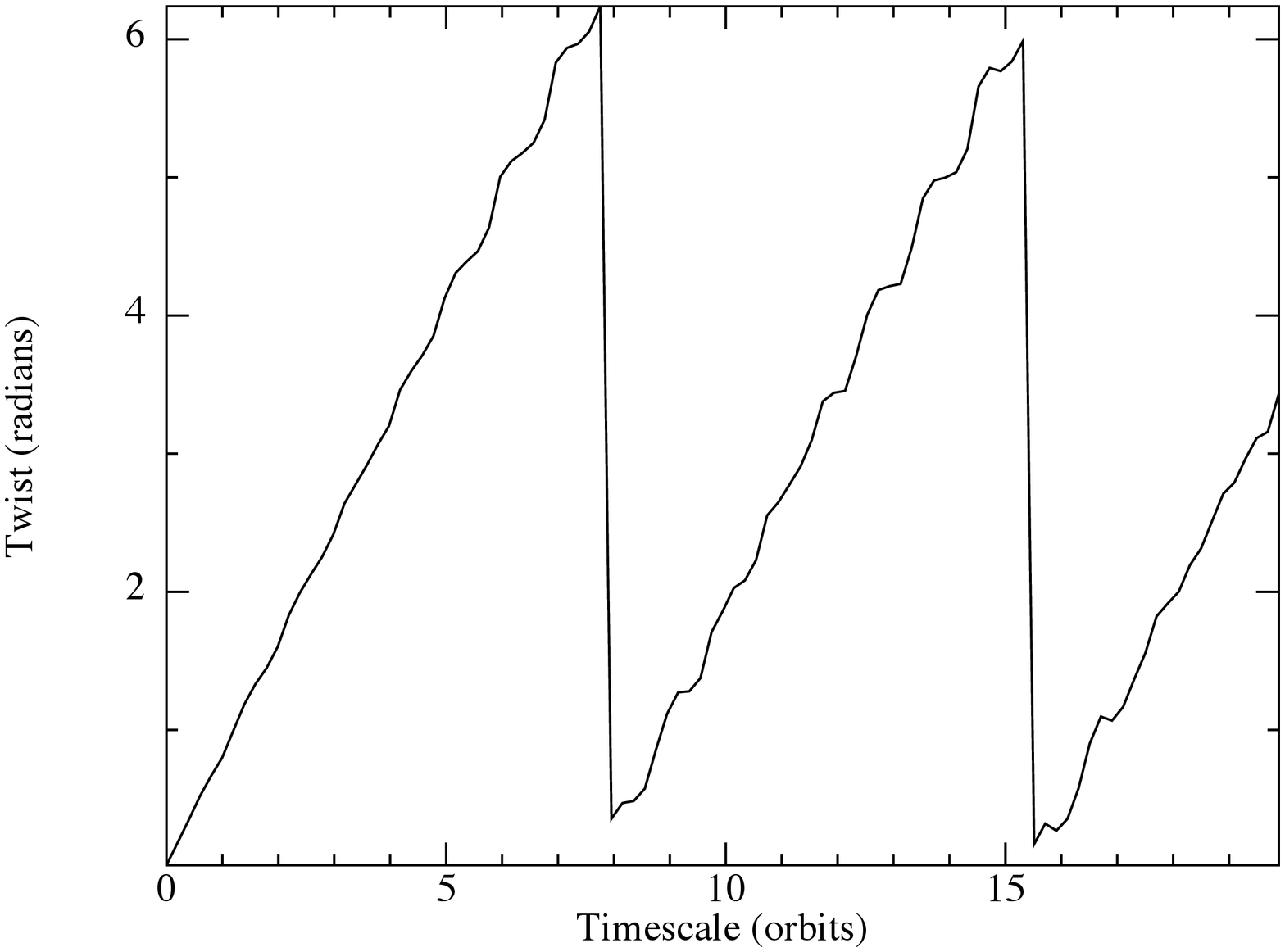}
\caption{An example of the description of twist in the disc as a function of time at a given radius. Here the time is plotted in orbits measured at the inner edge.}
\label{fig:twist_example}
\end{figure}

Here we outline how the precession in the disc was measured from the simulations and show some example results. In order to analyse the properties of the discs from the simulations, we discretise the disc into a set of thin spherical annuli and average the properties of interest across the particles in each of these bins. This process is described in detail in Section 3.2.6 of \citet{lodato_2010}. The twist, $\gamma(R)$, in our disc at a given radius is found by considering the unit angular momentum vectors at each radius bin in the disc.  With $l_x(R)$, $l_y(R)$ and $l_z(R)$ being the unit vectors in the Cartesian coordinate system, we assign
\begin{equation}
\gamma(R) = \tan^{-1}\left(\frac{l_y(R)}{l_x(R)}\right), \label{equation:gamma}
\end{equation}
for each radial annulus. This is repeated at every time step, so that we have a description of the twist as a function of time at each radial bin in our simulations. An example of the twist in this format is in Figure~\ref{fig:twist_example}. Here the twist is increasing in the disc when the gradient is positive. As the disc twists through a full $2\pi$ radians, the twist then jumps back to zero because Equation~\ref{equation:gamma} does not take into account the cumulative twist angle.

The precession time can be measured from Figure~\ref{fig:twist_example} directly by recording how long it takes for the disc to twist all the way around, equivalent to finding when the twist drops back to zero. This can be approximated by working out the gradient of the twist as a function of time and then using it to calculate the precession time in the disc. This is equivalent to calculating
\begin{equation}
t_{\rm p} = 2\pi \left(\frac{d\gamma}{dt}\right)^{-1}.
\end{equation}
Because this calculation has been done at each radial annulus, we now have the precession time as a function of the radius in the disc, averaged over the length of the simulation. An example of this was shown in Figure~\ref{fig:prec_example}.

The above analysis does not take the inclination of the disc into account. This disc was inclined at $30^{\circ}$, but this angle did not come into our expression for $t_p$ or explicitly in our analysis from the twist. As outlined in the derivation by \citet{Larwood:1996fk}, the Lense-Thirring precession is independent of the inclination between the disc and the black hole spin. Repeating the above analysis with discs at other angles confirms this.

\bibliographystyle{mn2e} 
\bibliography{master}

\label{lastpage}
\end{document}